\documentclass[a4paper,12pt]{article}
\usepackage{amssymb,amsmath, amsthm}
{

      \newtheorem{assumption}{Assumption}
  }
\usepackage{natbib}
\usepackage{pdflscape}
\usepackage{mathtools}
\usepackage{amssymb}
\usepackage{comment}
\usepackage{graphicx}
\usepackage{subcaption}
\usepackage[colorlinks,  citecolor={blue}]{hyperref}
\usepackage{float}
\usepackage{amsmath}
\usepackage[english]{babel}
\usepackage{amsfonts}
\usepackage{rotating}
\usepackage{epstopdf}
\usepackage{lscape}
\usepackage{pdflscape}
\usepackage{anysize}
\usepackage[official]{eurosym}
\usepackage{longtable}
\usepackage[a4paper,left=2.5cm,right=2.5cm,top=3cm,bottom=3cm]{geometry}
\usepackage{xspace}
\usepackage{array}
\usepackage{latexsym}
\usepackage{amsthm}
\usepackage{longtable}
\usepackage{setspace}
\usepackage{xspace}
\usepackage{authblk}
\usepackage[T1]{fontenc}
\usepackage[utf8]{inputenc}
\usepackage[pdftex]{color}

\usepackage{chngcntr}
\usepackage[utf8]{inputenc}
\usepackage{pdfsync}
\usepackage[toc,page]{appendix}
\usepackage{titlesec}
\usepackage[a4paper,bottom=3cm,left=2.5cm,right=2.5cm]{geometry}
\usepackage{graphicx}
\usepackage{multirow}
\usepackage[table]{xcolor}

\newcommand{\tx}[1]{\mbox{\;{#1}\;}}

\numberwithin{equation}{section}
\numberwithin{lemma}{section}
\numberwithin{theorem}{section}
\numberwithin{definition}{section}
\numberwithin{remark}{section}
\numberwithin{corollary}{section}
\numberwithin{proposition}{section}
\titleformat{\paragraph}
{\normalfont\normalsize\bfseries}{\theparagraph}{1em}{}
\titlespacing*{\paragraph}
{0pt}{3.25ex plus 1ex minus .2ex}{1.5ex plus .2ex}

\begin{document}
\begin{center}
{\LARGE  Mediation Analysis  Synthetic Control\let\thefootnote\relax\footnotetext{We have benefited from comments by Simone De Angelis and participants at several seminars, workshops, and conferences. The views expressed in this paper are those of the authors and do not necessarily reflect those of the Bank of Italy. Addresses for correspondence: Giovanni Mellace (giome@sam.sdu.dk) and Alessandra Pasquini (alessandra.pasquini@bancaditalia.it).}}\bigskip

{Giovanni Mellace* and Alessandra Pasquini**}

{*University of Southern Denmark; **Bank of Italy}

\today
\end{center}
\bigskip\bigskip\bigskip
 {\small \noindent \textbf{Abstract:} The synthetic control method (SCM) allows estimating the causal effect of an intervention in settings where panel data on a small number of treated and control units are available. We show that the existing SCM, as well as its extensions, can be easily modified to estimate how much of the ``total'' effect goes through observed causal channels. Our new mediation analysis synthetic control (MASC) method requires additional assumptions that are arguably mild in many settings. We illustrate the implementation of MASC in an empirical application estimating the direct and indirect effects of an anti-smoking intervention (California's Proposition 99).


}


{\small \renewcommand{\thefootnote}{\arabic{footnote}} %
\setcounter{footnote}{0} \pagebreak \setcounter{footnote}{0} \pagebreak %
\setcounter{page}{1} }

\section{Introduction}
The Synthetic Control Method (SCM), introduced by \cite{Aba2003} and further developed in \cite{Aba2010,Aba2015}, is becoming very popular in program evaluation. SCM is attractive, as it allows estimating the causal effect of an intervention even when data on only one treated and a few control units are available. This is possible by using information on the pre-intervention period to construct a ``synthetic control'', which mimics what would have happened to the treated unit in the post-intervention period in the absence of the intervention. \cite{Gob2016} compare SCM to other interactive fixed-effects models and find that it performs very well as soon as in post-intervention periods, the counterfactual outcome of the treated unit lies in the convex hull of the outcomes of the control units.  In a recent paper, \cite{Xu2017} further exploits the connection between SCM and interactive fixed-effect models and proposes a new method that combines both. Similarly, \cite{Ark2019} combine the SCM with another panel data method, i.e., the differences-in-differences approach, introducing the Synthetic Difference In Differences (SDID). \cite{Dou2016} propose a modification of SCM where the weights are not constrained to be positive and do not necessarily add up to one. \cite{Ben2018} further relax weight constraints and  correct for possible covariates imbalances, demonstrating that the synthetic control method can be seen as an inverse propensity score-weighting estimator. \cite{Gun2021} proposes a modification of SCM that allows to estimate the entire counterfactual distribution. \cite{Amj2018} introduce the Robust Synthetic Control (RSC), which allows to deal with missing data and de-noises the outcome under a different algorithm. \cite{Amj2019} propose a modification of the RSC which allows for the presence of covariates. Finally, \cite{Ath2017} propose a new method that includes synthetic control and other panel data methods as a special case. 

Although all of those methods are very well-suited for estimating the ``total'' effect of an intervention, they are mostly uninformative about the causal mechanisms that generated this effect. Often an intervention may first have  an impact on an intermediate outcome (hereafter also referred to as the ``mediator''), which affects the final outcome. In the presence of a mediator, the total effect can  generally be decomposed into a direct effect of the intervention and an indirect one generated through the mediator. Ignoring the presence of such intermediate outcomes might weaken the ability to draw a policy conclusion. For example, if the direct and indirect effects are both large and similar in magnitude but with opposite signs, merely looking at the total effect may fail to provide a complete picture of the consequences of the policy. Moreover, it is often important to quantify the direct and indirect effects to better target the intervention. Consider the huge decrease in tobacco consumption after the introduction of California's anti-tobacco law, Proposition 99,  estimated in \cite{Aba2010}. Proposition 99 not only increased the tobacco price but also introduced several anti-tobacco informational campaigns. It would be extremely relevant for a policy-maker to know how much of the decrease in tobacco consumption triggered by Proposition 99 is due to a price increase and investment in informational campaigns respectively. 


Mediation analysis is a standard approach to dealing with these kind of issues. The main challenge in mediation analysis is that the identification of the direct and indirect effects requires knowledge about the potential outcome an individual would get if the potential mediator was set to the value it would have taken under the opposite treatment status than the one observed. This is never observed for any individual. A large part of the literature focuses on identification and estimation of direct and indirect effects under sequential conditional independence (see \citealt{Pea2001,Rob2003,Ima2010,Ima2013,Van2012,Hub2013,Van2012,Hub2016,Hub2017}). The idea behind this approach is that once we check for observed characteristics and the conditionally independent treatment, the potential outcomes are independent of the potential mediators. To the best of our knowledge, none of the existing methods are specifically designed for panel data  and cannot directly be applied in settings with one or a few treated units and a few control units.\footnote{One exception is \cite{Deu2018}. However, they consider a framework with a randomized intervention with non-perfect compliance.} 
   
This motivates the introduction of  our mediation analysis synthetic control (MASC) method, a generalization of the SCM that allows decomposing the total effect of an intervention into its direct and indirect components. Given that SCM already requires all unobserved confounders either to be time invariant or to affect treated units and control units identically, the only additional assumption required by MASC is that we are able to find control units with mediator  values similar to those of the treated unit after treatment. This is an advantage with respect to the cross-sectional framework, where the additional assumptions required  to identify the direct and indirect effects are usually much stronger than the one needed to identify the total effect.  Moreover, the plausibility of the assumptions required for MASC can be judged by checking the overlap in pre-treatment outcomes and in covariate values similarly to a standard SCM. 

MASC can be easily implemented by using existing SCM algorithms and any of the aforementioned new extensions (with the exception of \citealt{Ark2019}). Indeed, as we will discuss in more detail below, to identify the direct and indirect effects, MASC re-weights control unit post-intervention outcomes by choosing weights that minimize the distance between treated and synthetic unit in pre-intervention observable characteristics (including pre-intervention values of the outcome and the mediator) as well as in post-intervention values of the mediator. Intuitively, this allows us to mimic what would have happened to the treated unit in absence of the intervention if the mediator value were set to the potential mediator under treatment. As we mentioned above, this is the main challenge of mediation analysis. Following \cite{Aba2010}, we illustrate MASC with a simple dynamic factor model with interactive fixed effects and show that both the direct and the indirect effect estimators are unbiased as the number of pre-intervention periods goes to infinity.

The rest of the paper is organized as follows: Section \ref{theor} introduces MASC; Section \ref{inf} proposes possible inference procedures; Section \ref{app} includes an empirical application to the introduction of Proposition 99; and Section \ref{con} concludes. All the technical proofs are relegated to the online appendix.

\section{The Mediation Analysis Synthetic Control Method} \label{theor}
\subsection{Notation and theoretical framework}\label{note}
Let us assume that we are interested in the effect of an intervention, $D$, implemented at time $T$, on an outcome, $Y$. Suppose that part of the effect of $D$ on $Y$ goes through an observed intermediate outcome (mediator), $M$.  The total effect of the intervention on the final outcome can be decomposed into an indirect effect, which goes through $M$, and a residual effect, commonly known as the ``direct effect'', which could also go through other (possibly unobserved) causal pathways. Although often crucial for policy conclusions, identification of the direct and the indirect effects may be challenging. 
To see this, let $D_{it}$ be a binary indicator that is equal to one if  unit $i$ is exposed to the intervention at time $t$. We will refer to units that are exposed to the intervention as ``treated'' and to those that are not exposed as ``control''. Using the potential outcome framework (see, e.g., \citealt{Rub74}) for each unit, $i$, we can define the potential mediator at time $t$ as follows:
\begin{eqnarray*}
M_{it}(d) \tx{for d}\in\{0,1\}.
\end{eqnarray*}
$M_{it}(d)$ is the value that the mediator of unit $i$ would take, at time t, if $D_{it}$ is set to $d$. Assuming that there are no anticipation effects on the mediator in the pre-intervention period and that the standard stable unit treatment value assumption (SUTVA) holds, the observed and the potential mediators are related through the following observation rule:
\begin{eqnarray*}
M_{it}=M_{it}(0)(1-D_{it})+M_{it}(1)D_{it}.
\end{eqnarray*}
Note that $M_{it}$ is always equal to $M_{it}(0)$ for both treated unit and control units in the pre-intervention period, $t<T$,  and that we can observe only one of the two potential mediators for each unit in the post-intervention period, $t\geq T$. 

Similarly, for each unit $i$ at time $t$, we define  the potential outcomes as 
\begin{eqnarray*}
Y_{it}(d, M_{it}(d'))\equiv Y_{it}^{d, M_{d'}} \tx{for} d, d' \in\{0,1\}.
\end{eqnarray*}
$Y_{it}^{d, M_{d'}}$ is the value that the outcome of unit $i$ would take at time $t$ if we set $D_{it}=d$ and  $M_{it}=M_{it}(d')$.  The potential outcome is a function of both the treatment  and the potential mediator. We assume that the potential mediator is only a function of the treatment and unaffected by future values of the outcomes \footnote{We do allow the potential mediator to depend on past values of the outcome.}. Under SUTVA, and assuming no anticipation effects in the pre-intervention period, the observed and the potential outcomes are related by the following observation rule:
\begin{eqnarray*}
Y_{it}=Y_{it}^{0,M_{0}}(1-D_{it})+Y_{it}^{1, M_{1}}D_{it}.
\end{eqnarray*}
Differently from the standard setting, each unit has four instead of two potential outcomes. As usual, only one potential outcome between $Y_{it}^{0,M_{0}}$ and $Y_{it}^{1,M_{1}}$ can be observed per unit in each period, while $Y_{it}^{0,M_{1}}$ and  $Y_{it}^{1,M_{0}}$ are never observed for any unit in any period. Assuming no anticipation effects, $Y_{it}=Y_{it}^{0,M_{0}}$  in the pre-intervention period  for all units.  

Following the synthetic control literature, we will define our parameters of interest with respect to a single treated unit. This contrasts with the standard mediation analysis literature, in which the total, direct, and indirect effects are defined as averages, either with respect to the whole sample (\citealt{Pea2001,Rob2003,Ima2010,Ima2013,Van2012,Hub2013}) or with respect to the treated units (\citealt{Van2012,Hub2017}). Nonetheless, if more than one unit is exposed to the intervention (see \citealt{Gob2016,Adh2015}) our method can be easily used to decompose the average treatment effect on the treated. 

We assume that we observe $J$ units ordered such that units 1 through $n$ are treated, while units $n+1$ through $J$ are controls. Without loss of generality, we will present our results for the first treated unit (unit 1) only. Since we have four potential outcomes instead of two, we can now define more parameters than in the standard synthetic control framework, each measuring the effect implied by a different thought experiment. Indeed, each potential outcome represents a different state of the world, and one can in principle define effects by calculating the difference between a pair of potential outcomes. Intuitively, $Y_{1t}^{0,M_{0}}$ and $Y_{1t}^{1,M_{1}}$ measure the value that the outcome of the first treated unit would take with and without intervention. On the other hand, $Y_{1t}^{0,M_{1}}$ and $Y_{1t}^{1,M_{0}}$  measure the values that the outcome of the treated unit would take if the value of the mediator were pushed to the value it would take under the opposite treatment status. Intuitively, given that policy-makers typically  cannot choose which value  the mediator takes under treatment, $Y_{1t}^{0,M_{1}}$   is arguably a more interesting counterfactual than  $Y_{1t}^{1,M_{0}}$. Indeed, it is often easier to push $M$ to $M(1)$ in the absence of the intervention, e.g., by implementing alternative policies that target the mediator directly, than $M$ to $M(0)$ with the intervention, as this would require ``neutralizing'' the effect of the intervention on the mediator and therefore implementing additional policies that have the opposite effects on the mediator. For this reason and since constructing a synthetic  $Y_{1t}^{1,M_{0}}$ requires stronger assumptions, as well as the ability to observe more than one treated unit, we will only focus on parameters that require knowledge of $Y_{1t}^{0,M_{1}}$ and relegate to Appendix \ref{indirect} our identification results for parameters that include $Y_{1t}^{1,M_{0}}$. 


The effects  of interest with regard to unit 1 are the total effect, $\alpha_{1t}$, which compares the outcomes the treated unit would get with and without the intervention; the direct effect, $\theta_{1t}(M_{1t}(1))$, which compares the treated unit potential outcome \textit{with} the intervention and the outcome  \textit{without} intervention when the mediator is set to the value it would have taken with the intervention; and the indirect effect, $\delta_{1t}(0)$, which measures the effect of pushing the mediator to its level under the intervention but without implementing the intervention. All parameters are assumed to be zero in the pre-intervention period. In the post-intervention period they are defined as
\begin{eqnarray*}\label{eq3}
\alpha_{1t}&=&Y_{1t}^{1,M_{1}}-Y_{1t}^{0,M_{0}},\\
\theta_{1t}(M_{1t}(1))&=&Y_{1t}^{1,M_{1}}-Y_{1t}^{0,M_{1}},\\
\delta_{1t}(0)&=&Y_{1t}^{0,M_{1}}-Y_{1t}^{0,M_{0}}, \quad t\geq T.
\end{eqnarray*} 
It is easy to see that the total effect, $\alpha_{1t}$, can be decomposed into\footnote{In the mediation literature the following alternative decomposition  is often also considered:
\begin{eqnarray*}
\alpha_{1t}&=&Y_{1t}^{1, M_{1}}-Y_{1t}^{0,M_{0}}, \\&=& Y_{1t}^{1, M_{1}}  - Y_{1t}^{1, M_{0}} + Y_{1t}^{1, M_{0}} - Y_{1t}^{0, M_{0}},\\ &=& \delta_{1t}(1) + \theta_{1t}(M_{1t}(0)), \\
\end{eqnarray*} 
where $\delta_{1t}(1)=Y_{1t}^{1,M_{1}}-Y_{1t}^{1,M_{0}}$ and $\theta_{1t}(M_{1t}(0))=Y_{1t}^{1,M_{0}}-Y_{1t}^{0,M_{0}}$. We decided not to focus on this decomposition for two reasons. First, as we argue above, a policy-maker would need to be able to neutralize the effect of the treatment on the mediator to reproduce  $Y_{1t}^{1,M_{0}}$. Second, identification of $Y_{1t}^{1,M_{0}}$ requires additional assumptions and the ability to observe multiple treated units.  
}:
\begin{eqnarray*} 
\alpha_{1t}&=&Y_{1t}^{1, M_{1}}-Y_{1t}^{0,M_{0}},\\&=& Y_{1t}^{1, M_{1}} - Y_{1t}^{0, M_{1}}+ Y_{1t}^{0, M_{1}} - Y_{1t}^{0, M_{0}},\\
&=&\theta_{1t}(M_{1t}(1)) + \delta(0).
\end{eqnarray*}
The decomposition above shows that if $\alpha_{1t}$ is identified, identifying $\theta_{1t}(M_{1t}(1))$ automatically implies identification of $\delta_{it}(0)=\alpha_{1t}-\theta_{1t}(M_{1t}(1))$.

To identify the total effect, SCM takes advantage of the panel structure and uses the pretreatment period to create a ``synthetic control'' unit that is similar to the treated unit both in terms of observables and in terms of unobservables. The latter need to be somehow stable between the pre-and post-intervention period. For example, the factor model described in \cite{Aba2010} implicitly assumes that all the  unobserved confounders are either time invariant or that they change in the same way for treated unit and control units.   The synthetic control mimics what would have happened to the treated unit in the post-intervention period in the absence of the intervention. In other words, SCM creates a synthetic value of $Y_{1t}^{0,M_{0}}$ in the post-intervention period. The synthetic control is built by using a linear combination of the control units. Under the assumption that all unobservables are either time invariant or affect the outcome of treated unit and control units in the same way, the  synthetic control can be built by a weighted average of the post-treatment outcomes of the control units using weights that are chosen to minimize the distance between the pre-intervention observable characteristics (including pre-intervention outcomes) of the treated unit and synthetic units (see \citealt{Aba2010}). Notice that for the SCM to work $Y_{1t}^{0,M_{0}}$ has to lie in (be close to)  the convex hull of the non-treated units post-intervention outcomes. 

Most of the methods used in mediation analysis rely on the so called sequential conditional independence assumption (SCIA). Under this assumption, the total effect, direct, and indirect effects can be recovered using control units and controlling for observable characteristics. When there is a single treated unit, however, we cannot invoke the SCIA. To overcome this issue, our MASC uses the SCM idea to create  a ``synthetic'' value of $Y_{1t}^{0, M_{1}}$ in the post-intervention period. To reconstruct $Y_{1t}^{0, M_{1}}$, we propose to use a linear combination of the control unit post-intervention outcomes, choosing weights that minimize the distance between treated unit and control units pre-intervention observable characteristics as well as post-intervention values of the mediator. The intuition is that choosing the weights that minimize the distance between the treated unit and synthetic control with respect to post-treatment values of the mediator as well will mimic what would have happened to the treated unit in the absence of the intervention but fixing the mediator value to its value in presence of the intervention. The additional assumption required by MASC is that $Y_{1t}^{0, M_{1}}$ has to lie in (be close to) the convex hull of the non-treated units post-intervention outcomes.  This is in contrast to the cross-section mediation analysis literature, in which the additional assumptions required to decompose the total effect are far stronger than the one required to identify it. 

To better illustrate our approach, in the spirit of  \cite{Aba2010}, we will introduce a factor model and show that, under this model, the bias is bounded by a measure which goes to zero as the number of pre-intervention periods goes to infinity. In our factor model, the potential mediators of unit $i$ are given by
\begin{eqnarray*}
M_{it}(d)&=&\gamma_t+\beta_tZ_i+\vartheta_t\varrho_i+\psi_t d+\nu_{it}, 
\end{eqnarray*}
where $\gamma_t$ is an unknown common factor with constant factor loadings across units. $Z_i$ is a $(p \times 1)$ vector of observed covariates;  $\beta_t$ is a $(1 \times p)$ vector of unknown parameters; $\vartheta_t$ is a $(1 \times v)$ vector of unobserved common factors; $\varrho_i$ is a $(v \times 1)$ vector of unknown factor loadings; $\psi_{it}$ is an unknown parameter describing the impact of the treatment on the mediator; and $\nu_{it}$ are unobserved transitory shocks. Notice that our model does not impose any extra restrictions on the unobservables that are only allowed to change over time in the same way for all units. It is easy to see that if this assumption is violated for the mediator, SCM would not be able to recover $Y_{1t}^{0, M_{0}}$, as $M$ would be a post-treatment time-varying confounder, i.e., the assumptions of a standard SCM would also be violated.  

Similarly, we assume that the four potential outcomes are given by
\begin{eqnarray*}
Y_{it}^{d,M_{d'}}&=&\zeta_t + \eta_tX_i+\lambda_t\mu_i+\varphi_t(1)M_{it}(d')+\rho_t(M_{it}(d'))\cdot d+\epsilon_{it}
\end{eqnarray*}
where $\zeta_t$ is an unknown common factor with constant factor loadings across units; $X_i$ is an $(r \times 1)$ vector of observed covariates that includes all the variables included in $Z_i$, but might also include other observable variables, which affects the treatment and the outcome but not the mediator; $\eta_t$ is a $(1 \times r)$ vector of unknown parameters; $\lambda_t$ is a $(1 \times F)$ vector of unobserved common factors; $\mu_i$ is an $(F \times 1)$ vector of unknown factor loadings; $\epsilon_{it}$ are unobserved transitory shocks; and  $\varphi_{it}(d)$ and $\rho_{it}(M_{it}(d))$ capture the impact on the potential outcomes of the potential mediator and the treatment, respectively. In this model, the total, direct, and indirect effects of unit 1 are then given by
\begin{eqnarray*}
\alpha_{1t}&=&\varphi_t(1)M_{1t}(1)-\varphi_t(0)M_{1t}(0)+\rho_t(M_{1t}(1)), \\
\theta_{1t}(M_{1t}(1))&=&\rho_t(M_{1t}(1))+(\varphi_t(1)-\varphi_t(0))M_{1t}(1), \\
\delta_{1t}(0)&=&\varphi_t(0)(M_{1t}(1)-M_{1t}(0)).
\end{eqnarray*}
Notice that the unobservables enter the potential outcomes equations in exactly the same way as in \cite{Aba2010}; thus, no extra restrictions are imposed on them.

Therefore, as also mentioned above, to estimate the total effect we can just use a standard SCM. 
In particular, we assume that there exists a $(1 \times (J-n))$ vector of  weights $L^*=(l_{n+1}^*,...,l_J^*)$ that are positive, adding up to 1, and such that in the post-intervention period $$Y_{1t}^{0,M_{0}}=\sum_{i=n+1}^J l^*_iY_{it}.$$ 
As in \cite{Aba2015}, we assume that $\forall \ t=1,...,T-1$, $L^*$ also satisfies
\begin{eqnarray*}
\sum_{j=n+1}^{J}l_j^*Y_{jt}&=& Y_{1t},\\
\sum_{j=n+1}^{J}l_j^*M_{jt}&=&M_{1t}, \\ 
\sum_{j=n+1}^{J}l_j^*X_{j}&=&X_{1}.
\end{eqnarray*} 

This justifies the choice of weights that minimize the distance between the observable characteristics of the treated unit and the control units in the pre-treatment period. More formally, let $\Omega^{\alpha}_1=(X_1, Y_{11},\ldots,Y_{1,T-1},M_{11},\ldots,M_{1,T-1})$ be a $((2(T-1)+r) \times 1)$ vector, $\omega^{\alpha}_{0i}=(X_i, Y_{i1},\ldots,Y_{i,T-1},M_{i1},\ldots,M_{i,T-1})$  be a $(1 \times (2(T-1)+r))$ vector, and $\Omega^{\alpha}_{0}=(\omega^{\alpha}_{0,n+1},\ldots,\omega^{\alpha}_{0J})'$. Then
\begin{eqnarray*}
L^*&=&\min_{l_{n+1},...,l_J} ||\Omega^{\alpha}_{1}-L\Omega^{\alpha}_{0}||\\
&s.t.& l_{n+1}\leq 0,...,l_J\leq0, \sum_{i=n+1}^J l_i=1,
\end{eqnarray*}
where $||\Omega^{\alpha}_{1}-L\Omega^{\alpha}_{0}||=\sqrt{\left(\Omega^{\alpha}_{1}-L\Omega^{\alpha}_{0}\right)'\left(\Omega^{\alpha}_{1}-L\Omega^{\alpha}_{0}\right)}$.
It is also possible to give more weight to specific observable characteristics by using the alternative distance $||\Omega^{\alpha}_{1}-L\Omega^{\alpha}_{0}||_V=\sqrt{\left(\Omega^{\alpha}_{1}-L\Omega^{\alpha}_{0}\right)'V\left(\Omega^{\alpha}_{1}-L\Omega^{\alpha}_{0}\right)}$ (see \citealt{Aba2010} for a data driven procedure to choose $V$). 

Let $\hat{Y}_{1t}^{0,M_{0}}=\sum_{i=n+1}^J l^*_iY_{it}$, \cite{Aba2010} show that if $L^*$ exists, for $t\geq T$ 
$$ 
E(\hat{Y}_{1t}^{0,M_{0}})=Y_{1t}^{0,M_{0}}+o(T)
$$
Consequently,  estimating the total effect as $\hat{\alpha}_{1t}=Y_{1t}-\hat{Y}_{1t}^{0,M_{0}}$ is justified by the fact that
\begin{equation}
\lim_{T\to \infty}E(\hat{\alpha}_{1t})=\alpha_{1t} \tx{ }\forall\tx{ } t \geq T
\end{equation}
The estimation of $Y_{1t}^{0, M_{1}}$, and thus the direct and indirect effects,  requires additional constraints but no extra assumptions on the unobservable in the potential outcomes equations. Our goal is to construct a ``synthetic'' unit which is identical to the treated unit, not affected by the intervention, and, at the same time, has the same value of the mediator as the treated unit. Similar to standard SCM, we want to find a $(1 \times (J-n))$ vector weights $W_t^*=(w_{n+1,t}^*,...,w_{Jt}^*)$ that are positive, adding up to 1, and such that in the post-intervention period $$Y_{1t}^{0,M_{1}}=\sum_{i=n+1}^J w^*_{it}Y_{it}.$$ Notice that, in our simple factor model, $Y_{1t}^{0, M_{1}}$ depends on the value that $M$ takes at time $t$ only.\footnote{It is easy to let $Y_{1t}^{0, M_{1}}$ depend on all the values that the mediator takes between $T$ and $t$. 
This is done by replacing $\Omega^{\theta_{t'}(1)}_1$ and $\omega^{\theta_{t'}(1)}_{0i}$ defined  below with  $\Omega^{\theta_{t'}(1)}_1=(X_1, Y_{11},\ldots,Y_{1,T-1},M_{11},\ldots,M_{1,T-1}, M_{1,T},\ldots,M_{1,{t'}})$ and $\omega^{\theta_{t'}(1)}_{0i}=(X_i, Y_{i1},\ldots,Y_{i1,T-1},M_{i1},\ldots,M_{i,T-1},M_{i,T},\ldots,M_{i,{t'}})$, respectively.} Also notice that the weights need to be calculated in each post-intervention period in this model. 
Let ${t'}\geq T$ be the time at which we want to estimate the direct effect. Similar to \cite{Aba2010} we assume that $W^*_{{t'}}$ exists and  satisfies $\forall \ t=1,...,T-1$:
\begin{eqnarray*}
\sum_{j=n+1}^{J}w_{j{t'}}^*Y_{jt}&=& Y_{1t},\\
\sum_{j=n+1}^{J}w_{j{t'}}^*X_{j}&=&X_{1},
\end{eqnarray*} 
and $\forall \ t=1,...,T-1,{t'}$, 
$$
\sum_{j=n+1}^{J}w_{j{t'}}^*M_{jt}=M_{1t}.  
$$

The vector of weights, $W_{{t'}}^*$, is then estimated in a similar way as $L^*$. The only difference is that we now need to include the post-treatment mediator in the distance. More formally, if we let $\Omega^{\theta_{t'}(1)}_1=(X_1, Y_{11},\ldots,Y_{1,T-1},M_{11},\ldots,M_{1,T-1}, M_{1,{t'}})$, $\omega^{\theta_{t'}(1)}_{0i}=(X_i, Y_{i1},\ldots,Y_{i,T-1},M_{i1},$ $\ldots,M_{i,T-1}, M_{i,{t'}})$, and $\Omega^{\theta_{t'}(1)}_{0}=(\omega^{\theta_{t'}(1)}_{n+1},\ldots,\omega^{\theta_{t'}(1)}_{J})'$, then
\begin{eqnarray*}
W_{t'}^*&=&\min_{w_{n+1,{t'}},...,w_{J{t'}}} ||\Omega^{\theta_{t'}(1)}_{1}-W_{t'} \Omega^{\theta_{t'}(1)}_{0}||_V\\
&s.t.& w_{n+1,{t'}}\leq 0,...,w_{J{t'}}\leq 0, \sum_{i=n+1}^J w_{i{t'}}=1,
\end{eqnarray*}
where $||\Omega^{\theta_{t'}(1)}_{1}-W_{t'} \Omega^{\theta_{t'}(1)}_{0}||_V=\sqrt{\left(\Omega^{\theta_{t'}(1)}_{1}-W_{t'} \Omega^{\theta_{t'}(1)}_{0}\right)'V\left(\Omega^{\theta_{t'}(1)}_{1}-W_{t'} \Omega^{\theta_{t'}(1)}_{0}\right)}$. Notice that we only have one mediator in the post-intervention period and several pre-intervention variables. Thus, we suggest to choose  $V$ such that equal weights are given to pre- and post- intervention information. 

Let $\hat{Y}_{1{t'}}^{0,M_{1}}=\sum_{i=n+1}^J w^*_{i{t'}}Y_{i{t'}}$, as we show in the appendix, if $W^*_{{t'}}$ exists, under standard regularity conditions:
$$ 
E(\hat{Y}_{1{t'}}^{0,M_{1}})=Y_{1{t'}}^{0,M_{1}}+o(T).
$$
This allows us to  estimate the direct effect as ${\theta}_{1{t'}}(M_{1t}(1))$ and the indirect effect as $\delta_{i{t'}}(0)$ since $$\hat{\theta}_{1t'}(M_{1{t'}}(1))=Y_{1{t'}}-\hat{Y}_{1{t'}}^{0,M_{1}}, \qquad \hat{\delta}_{1t'}(0)=\hat{\alpha}_{1{t'}}-\hat{\theta}_{1t'}(M_{1{t'}}(1)),$$ 
respectively, as it implies
\begin{eqnarray*}
\lim_{T\to \infty}E(\hat{\theta}_{1{t'}}(M_{1t}(1)_{1t})=\theta_{1{t'}}(M_{1t}(1)_{1t}) \tx{ }\forall\tx{ } t \geq T, \\
\lim_{T\to \infty}E(\hat{\delta}_{i{t'}}(0))=\delta_{i{t'}}(0) \tx{ }\forall\tx{ } t \geq T
\end{eqnarray*}

\subsection{Additional assumptions of MASC}

Intuitively, $W^*_{{t'}}$ only exists if, in addition to the assumptions needed for a standard SCM, there is also an overlap in the post-intervention values of the mediator.
This assumption basically requires that the post-intervention value of the mediator of the treated units is not an outlier in comparison to the one of the units in the donor pool. This assumption is, of course, not required to identify the total effect and in some applications might be violated, even though the standard SCM assumption are satisfied.    
However, similarly to the standard SCM, the plausibility of our additional assumption can be graphically assessed by looking at the overlap in the pre-intervention period between the observed outcome, $Y_{1{t}}$, and the synthetic outcome $\hat{Y}_{1{t}}^{0,M_{1}}$ together with the overlap in the post-intervention period between the mediator of the treated and those of the synthetic unit.

\subsection{Inference}\label{inf}
Inference can be carried over in a similar manner as for the standard synthetic control method. For example, one can run similar placebo tests as the one suggested in \cite{Aba2010}, \cite{Aba2015} and \cite{Aba2019}, estimating the effects (in our case also the direct and indirect effects) of the intervention either before its implementation or for units not exposed to it. Another possibility is to follow the approaches outlined in \cite{Che2018} and \cite{Gob2016}, to cite a few.
We follow \cite{Aba2010} and base our inference on the estimated placebo effect, excluding all the placebo units having a pre-treatment Root Mean Squared Prediction Error (RMSPE) which is more than $n$ times bigger than those of the treated units. For unit $i$ and synthetic outcome $\hat{Y}^{d,M_{d'}}$,  the pre-intervention RMSPE can be defined as  
\begin{eqnarray*}
RMSPE^{\hat{Y}^{d,M_{d'}},pre}_i=\frac{\sum_{t=1}^{T-1}(Y_{i,t}-\hat{Y}_{i,t}^{d,M_{d'}})^2 }{T-1}.
\end{eqnarray*}
The p-values can be calculated as the portion of selected placebo units whose total (direct or indirect) estimated effect is equal or bigger to those of the treated unit. As underlined in \cite{Aba2015}, in absence of randomization, the p-values can only be interpreted as the probability to obtain an effect at least as large as those of the treated units. \par

\section{A user-friendly empirical example}\label{app}
In this section we apply MASC to estimate how much of the effect of California's Proposition 99  on cigarettes consumption found in \citealt{Aba2010} can be attributed to the increase in  prices due to the excise tax.  We take a  user-friendly approach by describing step by step all the choices we need to make.  

\subsection{Institutional framework and data} \label{set}
Proposition 99 was a large-scale tobacco control program introduced in California in 1988 from a voter initiative. It introduced an excise tax of 25 cents per pack of cigarette. The revenues for the tax had to be split mainly between a tobacco prevention (80\%) and a cessation program (20\%) supporting  the health system for tobacco-related issues and tobacco-related research (\citealt{Bal1997}).  Similarly to previous studies (\citealt{Sie2002}), using the SCM, \cite{Aba2010} find strong negative effects on consumption. Although both the tax and the health education program sought to reduce smoking prevalence, they differ in many respects. The former generates revenues for the state, but might have smaller  long-term effects, while the latter might be more successful in the long run, but generates expenses. Therefore, it is crucial to determine to what extent each of its components contributed to the success of Proposition 99. \cite{Hu1995} used time-series monthly data to isolate the impact of the excise tax from those of the media campaign. They found that both had a negative effect on cigarette consumption, although with the excise tax having the largest. We apply MASC to provide further evidence on the role of the tax increase as a causal channel.

Following \cite{Aba2010}, we get data on per-capita cigarette consumption in packs (our outcome), on retail price per pack of cigarette and on the federal and state taxes from \cite{Orz2005} (the database is publicly available through the Centers for Disease Control and Prevention).  We use  is cigarette price per pack, measured as the sum of retail price and federal and state taxes as our mediator. The annual state-level database over the period 1970-2000 includes also other relevant control variables (for additional details, see the next section and \cite{Aba2010}).

\subsection{MASC implementation}\label{Impl}

\textbf{Selection of the donor pool.} As in SCM, MASC implementation requires a valid donor pool. It cannot include any unit that was affected by the intervention. Moreover, neither units in the donor pool nor treated units can experience large idiosyncratic exogenous shocks, affecting the outcome during the period under study. Finally, it is strictly recommended to only include units in the donor pool that are similar to the treated units. Therefore, we exclude from the donor pool all states that adopted large-scale tobacco control program during our sample period and, following \cite{Aba2010}, the District of Columbia as it is unlikely to have characteristics similar to California. The states that implemented  tobacco interventions are Massachusetts, Arizona, Oregon, and Florida. Moreover, following \cite{Aba2010}, when re-estimating the total effect of Proposition 99,  we exclude  all the states that increased tax excises by more than 50 cents. These are Alaska, Hawaii, Maryland, Michigan, New Jersey, New York and Washington. However, as those states did not complement the tax excise with a tobacco control program (\citealt{Ins1994}),  not only we can safely include them when we estimate the direct effect; their inclusion can help us to better find a synthetic unit whose mediator is equal to the one of the treated unit under treatment, i.e., $M(1)$. Our final donor pools included 38 states for the total effect estimation and 45 states for the direct effect estimation (see Appendix \ref{data}).  \\

\textbf{Selection of covariates.} Our goal is to create a synthetic unit whose outcome resembles as close as possible the one the treated unit would have had in the absence of the intervention. Therefore, following \citealt{Aba2010}, we select a set of variables that are predictors of the outcome of interest: real per capita GDP by state (chained at 1997 prices), the percentage of population aged 15-24 and beer consumption per capita. All these predictors were averaged over the period 1980-1988. In addition, as in \citealt{Aba2010}, we included the pre-treatment values of the outcome in 1975 and all years between 1980 and 1988. For the same years we also include pre-treatment values of the mediator (as well as the post-treatment years values of the mediator when estimating the direct effect; see the next paragraph). One can include some predictors of the mediator, provided that their inclusion does not deteriorate the pre-treatment fits of the outcome. More details on the applied covariates are available in Appendix \ref{data}. \\

\textbf{Mediator lag selection.} 
In Section \ref{theor}, we have assumed that all the effects happen instantaneously. Nonetheless, in empirical applications the treatment may have a delayed impact on the mediator and/or the mediator may have a delayed effect on the outcome. The timing of the effects affects the choice of the post-intervention value of the mediator we need to include among the predictors. Suppose that mediator at time $t'-\phi$ affects the outcome at time $t'$, with $\phi\geq 0$. Consequently, if we want to estimate the effect at time $t'$, we cannot include mediator values at any time $t > t'-\phi$.  The choice of $\phi$ depends on what economic theory suggests. In our application, it is reasonable to think that an increase in cigarette prices should immediately affect consumption (\citealt{Ins1994}); thus, we chose  $\phi=0$. To check the sensitivity of the results to this choice, we repeat the estimation with a one-year lag. The results (available from the authors upon request) do not change substantially. 

Notice that we re-estimate the weights separately for each post-intervention period $t'$. Therefore, once we have chosen $\phi$, we can either impose post-treatment constraints on all time periods between $T$ and $t'-\phi$ or only in period $t'-\phi$.  Including all periods allows to control for potential dynamic effects of the mediator but increases the complexity of the model and the chance of having a bad overlap in the post-treatment mediator. 
In our application we have chosen to impose constraints in the entire period (from $T$ to $t'-\phi$). Nevertheless, the overlap remains good for most of the post-intervention period (see Figure \ref{fig4} in Appendix \ref{med_con}). This means that the further we move from the starting period $T$, the higher the number of post-intervention constraints included.  \\

\textbf{Control variables weights.} We have to choose the weights that are assigned to each constrain (i.e., the matrix $V$). In the calculation of the total effect, we can select the matrix $V$ using a cross-validation method, as suggested in \cite{Aba2015}. Following this approach, the matrix is not uniquely identified (\citealt{Klo2018}). Therefore, in the placebo tests and in the leave-one-out robustness check, we employ the matrix selected in the first step. In the calculation of the direct effect, we have to make one additional choice. There is only one (or few) constraint(s) on the mediator in post-intervention period and several constraints on covariates in the pre-intervention period. Given that it is particularly important to have a good overlap in the post-treatment period mediator of our treated and synthetic control units,  we suggest choosing $V$ to ensure that a similar weight is given to the pre- and post-intervention constraints. In the application, we assign 3/4 of the total weights to pre-treatment constraints  using cross-validation as for the total effect and the remaining 1/4 equally across the post-treatment constraints. \\

\textbf{Inference.} 
For inference, we follow \cite{Aba2010}; i.e., we estimate the placebo effects for all the units in the donor pool. Later on, to make inference for the direct and the total effects, we drop the results for all the placebo units whose pre-treatment RMSPE is more than five times those of the treated unit. For the indirect effect, instead, we drop the results for all the placebo units that have a big RMSPE in the estimation of either the direct or the total effect estimation. Finally, we derive  the p-values from the placebo tests as proposed in \cite{Aba2015} (see section \ref{inf} for details). \\

\begin{figure}[h!]
\begin{subfigure}{0.5\textwidth}
\centering\captionsetup{width=.8\linewidth}
\includegraphics[width=0.9\linewidth, height=7cm]{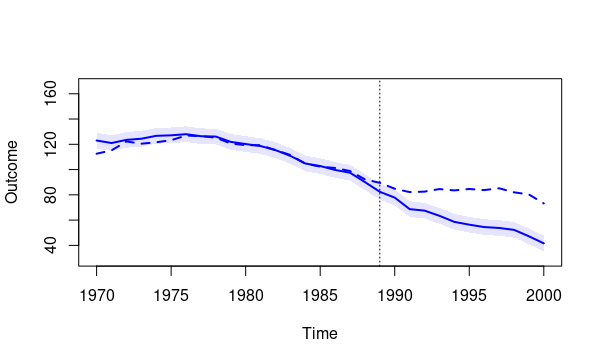} 
\caption{Total effect}
\label{fig1}
\end{subfigure}
\begin{subfigure}{0.5\textwidth}
\centering\captionsetup{width=.8\linewidth}
\includegraphics[width=0.9\linewidth, height=7cm]{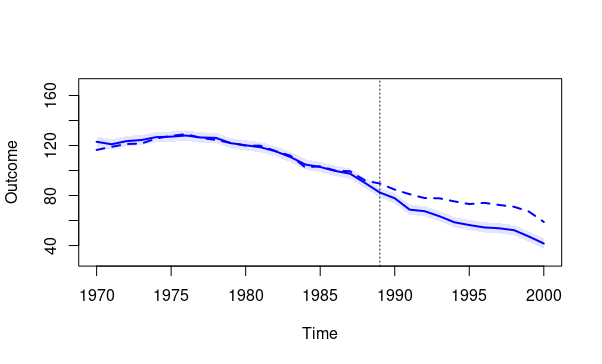} 
\caption{Direct effect}
\label{fig3}
\end{subfigure}
	\caption{Comparison of the outcome of the treated unit and those of the synthetic units. The solid line represents the treated unit outcome. The dashed line represents the synthetic unit outcome when $Y=Y^{0,M_{0}}$ (left) and when $Y=Y^{0,M_{1}}$ (right). The shaded area around the solid line corresponds to two standard deviations of the pre-treatment difference between the outcomes of the treated unit and those of the synthetic units. The vertical line corresponds to the first treatment period. In the direct effect estimation, different weights are used in each post-treatment period. For simplicity, we use those found in the last post-treatment period.}
\end{figure}

\textbf{Satisfaction of  MASC assumptions.}
First, we need to check that the overlaps in pre-treatment outcome between the treated unit and all synthetic units (the one used for estimating the total and the direct effects) are good enough. Figures \ref{fig1} and \ref{fig3} show that the overlaps for both the total and the direct effect are very good. To estimate the direct effect, it is important that there is a good overlap between the post-treatment values of the mediator. Figure \ref{fig4} in Appendix \ref{med_con} shows a good overlap in the post-treatment value of the mediator using weights obtained in the last estimation period, which is the more demanding, as it includes the highest number of post-intervention constraints (see the discussion about the selection of mediator lags).

Another requirement is that there are no anticipation effects of the intervention both on the outcome and on the mediator. As the policy was introduced in January 1989 after a ballot in November 1988, anticipation effects do not appear to be a major concern. 

We cannot have any spillover effects on the units in the donor pool. \cite{Aba2010} enumerates multiple ways in which this requirement maybe violated in our context. First, the tobacco industry may have diverted funds for campaigns or lobbying from other states to California. Nonetheless, if anything, such a phenomenon would downward bias the treatment effect estimation. Second, the anti-tobacco sentiment may have spread from California to neighboring states. Third, cross-border purchase from neighboring jurisdictions may increase after the introduction of the policy. To check for the last two potential spillovers, we check the robustness of our results to the exclusion of neighboring states from the donor pool, i.e., we repeat the estimation excluding Nevada (the other two neighboring states, Oregon and Arizona, were not in the donor pool). As shown in figure \ref{spill1} in the appendix, our results are fairly robust.  

Finally, there has to be no reverse effect of the outcome on the mediator. This requirement is typically less of a concern when it is assumed that the impact of the mediator on the outcome is not instantaneous (i.e., when $\phi>0$) as in many applications it is reasonable to assume that the outcome at time $t$ does not have an impact on the mediator at time $t-\phi$. As we set $\phi=0$, the satisfaction of this requirement is not as straightforward. We have to assume that cigarette prices do not instantaneously change to reflect the new consumption level. If there was perfect competition, the decrease in demand would immediately imply a decrease in cigarette prices. However, the empirical evidence suggests that the tobacco industry in the USA looks more like an oligopoly, and when Proposition 99 was implemented, prices did not seem to adjust quickly to the demand shock. The tax induced an increase in total cigarette prices amounting to 127\% of the tax increase (\citealt{Hu1995b}). 
Nevertheless, as a further check, we re-estimated the direct effect using the state tax as a mediator finding qualitatively similar results, which are available upon request.

\subsection{Results}\label{result}
Before discussing the results, we want to clarify how the direct and indirect effects should be interpreted. The direct effect captures the effect of the prevention and cessation program, those of the tobacco-related research activity (financed by the revenues of the taxes), and those of the stimulus that Proposition 99 gave to the development of local clean indoor air ordinance (\citealt{Sie2002}). The prevention and cessation program was present in state-wide media campaigns, community-based programs, and grants for the development of high school and community programs. In the first year, 82 million dollars was allocated to contribute to it \cite{Sie2002}. The proposition passed in November 1988 and it came into force in January 1989. Since the fiscal year in California goes from July to June, the funds for the campaign were already employed for half of 1989. Notice that as we estimate the direct effect $\theta_{1t}(M_{1t}(d))$ when $d=1$, we calculate the effect of these programs when accompanied by an increase in cigarette price. This may differ from the effect these programs would have had in its absence. 

Given that we estimate the indirect effect $\delta(d)$ when $d=0$, this is the effect that only an increase in cigarette prices would have had in the absence of other programs (i.e., the prevention and control program, the research activity, the increase in indoor clean air ordinances). As shown in Figure \ref{fig2} in Appendix \ref{med_con}, Proposition 99 induced a strong increase in cigarette prices since 1989 and remained stable up to 1996. After this year, the two prices started to converge until 1999, when the difference between the two prices sharply diverged again. 

\begin{figure}[h]
\centering\captionsetup{width=.7\linewidth}
\includegraphics[width=0.9\linewidth, height=11cm]{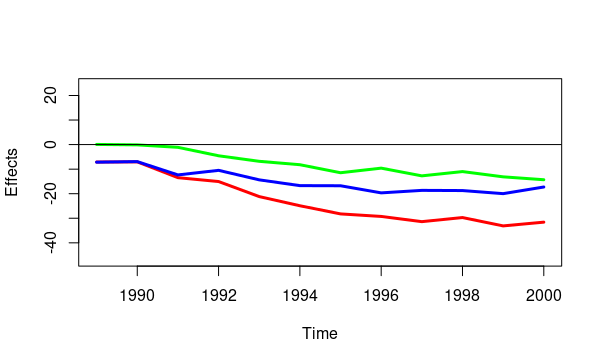} 
	\caption{Effects of Proposition 99. Total (red), direct (blue), and indirect (green) effects evolution over time.}
	\label{fig5}
\end{figure}

In Table \ref{tab2} and Figure \ref{fig5}, we report the estimates for the total, direct, and indirect effects\footnote{The results are robust to leaving one state out from the donor pool
(see Appendix \ref{spill}).}. 

\newgeometry{a4paper, left=2cm, right=2cm, top=2cm, bottom=2cm, nohead}
\begin{table}[H]
\centering
\caption{Effects of Proposition 99}\label{tab2}
{\footnotesize\begin{tabular}{lccc} \hline
Year & Total & Direct & Indirect \\ \hline
1989 & -7.1 & -7.17 & 0.06 \\
P-value & (0.22) & (0.31) & (1) \\
1990 & -7.04 & -6.93 & -0.11 \\
P-value & (0.3) & (0.19) & (1) \\
1991 & -13.47* & -12.35 & -1.12 \\
P-value & (0.07) & (0.13) & (0.75) \\
1992 & -15.05*** & -10.49 & -4.56 \\
P-value & (0) & (0.25) & (0.38) \\
1993 & -21.18** & -14.36*** & -6.81 \\
P-value & (0.04) & (0) & (0.19) \\
1994 & -24.92*** & -16.72*** & -8.2 \\
P-value & (0) & (0) & (0.13) \\
1995 & -28.23*** & -16.77 & -11.46*** \\
P-value & (0) & (0.13) & (0) \\
1996 & -29.26*** & -19.67*** & -9.6* \\
P-value & (0) & (0) & (0.06) \\
1997 & -31.38*** & -18.65*** & -12.73*** \\
P-value & (0) & (0) & (0) \\
1998 & -29.72*** & -18.72*** & -10.99 \\
P-value & (0) & (0) & (0.13) \\
1999 & -33.12*** & -19.98* & -13.13*** \\
P-value & (0) & (0.06) & (0) \\
2000 & -31.59** & -17.28 & -14.31*** \\
P-value & (0.04) & (0.13) & (0) \\
\hline
\end{tabular}}
\caption{P-values are displayed in brackets. * Significant at 10\%. ** Significant at 5\%. *** Significant at 1\%.}
\end{table}
\restoregeometry


In line with those of \cite{Aba2010}, our results show that Proposition 99 reduced the cigarette per-capita consumption from a minimum of 7 to a maximum of 33 packs each year. The direct effect, which is visible from the very first year of treatment, seems to have contributed the most to this decrease, although the contribution of the indirect effect increases over time. Notice that the total effect is significantly different from zero since 1991, the direct effect since 1993, and the indirect effect only in the second half of the period. The direct effect decreased slightly in 1992. This is explained by the fact that in 1992, Governor Pete Wilson suspended the media campaign that was  re-established  the next year, following a lawsuit by the American Lung Association. The fact that the decrease in 1992 is only visible in the direct effect while the indirect effect in unaffected, increases our confidence that we are correctly decomposing the total effect. 

Our results suggest that the contribution of the direct effect is higher than what was found by \cite{Hu1995} for the media campaign. However, we need to emphasize that our direct effect  incorporates not only the effect of the media campaign but also the effects of the other programs (i.e., community-based programs, grants and research activity). Our estimates of  the effects of the cigarette prices increase are also lower than in their estimation. Overall, our results suggest that the two policies combined were more successful in reducing cigarette consumption than only imposing a higher tax would have been.

\section{Conclusions}\label{con}
We introduced MASC, a new method that  combines the synthetic control method (\citealt{Aba2003,Aba2010,Aba2015}) with the mediation analysis approach and allows us to identify direct and indirect effects in a panel data frameworks with a low number of treated and control units. MASC is intuitive and easy to implement (i.e., publicly available SCM algorithms can be employed). Even though we show how to implement our method based on the original SCM algorithm of \cite{Aba2010,Aba2015}, one can alternatively use many of the new SCM-based methods such as \cite{Ath2017,Xu2017,Kre2016,Ben2018}; and \cite{Dou2016}. 
Finally, we provide an illustrative example where we decompose  the ``total'' effect of Proposition 99 on cigarette consumption, and we show that an increase in cigarette prices alone would have had a lower impact than in combination with the other programs. 

\section*{Appendix}
\appendix

\section{Derivation of ``Synthetic'' $Y_{1t}^{0M_{1}}$}\label{appY01}
To ease the notation, the subscript $t$ is dropped from the weights. 
Following \cite{Aba2010}, consider a generic vector of weights $W=(w_{n+1},...,w_J)'$ such that $w_j\geq0$ for all $j=n+1,...,J$ and $w_{n+1}+...+w_{J}=1.$ With these weights (and considering the factor model introduced in the text) the synthetic value of $Y_{1t}^{0M_{1}}$ is given by
\begin{eqnarray*}
\sum_{j=n+1}^{J}w_jY_{jt}=\zeta_t + \eta_t\sum_{j=n+1}^{J}w_jX_j+\lambda_t\sum_{j=n+1}^{J}w_j\mu_j+\varphi_t(0)\sum_{j=n+1}^{J}w_jM_{jt}(0)+\sum_{j=n+1}^{J}w_j\epsilon_{jt}.
\end{eqnarray*} 
The difference between the real potential outcome and the synthetic one is then 
\begin{eqnarray}
Y_{1t}^{0,M_{1}}-\sum_{j=n+1}^{J}w_jY_{jt}&=&\eta_t\left(X_1-\sum_{j=n+1}^{J}w_jX_j\right)+
\lambda_t\left(\mu_1-\sum_{j=n+1}^{J}w_j\mu_j\right)\notag\\&+&\varphi_t(0)\left(M_{1t}(I\{t\geq T\})-\sum_{j=n+1}^{J}w_jM_{jt}(0)\right)\notag\\&+&\sum_{j=n+1}^{J}w_j(\epsilon_{1t}-\epsilon_{jt}). \label{diff1}
\end{eqnarray}
  
Let $Y_i^P$ be the $((T-1) \times 1)$ vector with $t$th element equal to $Y_{it}$, $\epsilon_i^P$ the $((T-1) \times 1)$ vector with $t$th element equal to $\epsilon_{it}$, $\eta^P$ the $((T-1) \times r)$ matrix with $t$th row equal to $\eta_{t}$ and $\lambda^P$ the $((T-1) \times F)$ matrix with $t$th row equal to $\lambda_{t}$. Moreover, let $\varphi^P(0)$ be the $((T-1) \times 1)$ vector with $t$th element equal to $\varphi_{t}(0)$ and $M_i^P(0)$ the $((T-1) \times 1)$ vector with $t$th element equal to $M_{it}(0)$.  We can now write 
\begin{eqnarray*}
Y_{1}^{P}-\sum_{j=n+1}^{J}w_jY_{j}^P &=& \eta^P\left(X_1-\sum_{j=n+1}^{J}w_jX_j\right)+\lambda^P\left(\mu_1-\sum_{j=n+1}^{J}w_j\mu_j\right)\notag\\
&+& \varphi^P(0)\left(M_{1t}^P(0)-\sum_{j=n+1}^{J}w_jM_{jt}^P(0)\right)+\left(\epsilon_{1}^P-\sum_{j=n+1}^{J}w_j\epsilon_{j}^P\right). 
\end{eqnarray*} 
Note that we have $M_{1t}^P(0)$ as $t<T$. It is easy to see that:
\begin{eqnarray} 
\lambda^P\left(\mu_1-\sum_{j=n+1}^{J}w_j\mu_j\right) &=& Y_{1}^{P}-\sum_{j=n+1}^{J}w_jY_{j}^P- \eta^P\left(X_1-\sum_{j=n+1}^{J}w_jX_j\right)\notag\\
&-&\varphi^P(0)\left(M_{1t}^P(0)-\sum_{j=n+1}^{J}w_jM_{jt}^P(0)\right)\notag\\&-&\left(\epsilon_{1}^P-\sum_{j=n+1}^{J}w_j\epsilon_{j}^P\right)\label{A}
\end{eqnarray}
Similar to \cite{Aba2010}, assume that
\begin{assumption}\label{nonsing}
$\sum_{t=1}^{T-1}\lambda_t'\lambda_t$ is non-singular.
\end{assumption}
Assumption \ref{nonsing} is equivalent to assuming no perfect-collinearity among unobserved common factors and implies that $(\lambda^{P'}\lambda^P)^{-1}$ exists. We can then multiply both sides of  \ref{A} by $(\lambda^{P'}\lambda^P)^{-1}\lambda^{P'}$ to get
\begin{eqnarray*}
\mu_1-\sum_{j=n+1}^{J}w_j\mu_j &=&(\lambda^{P'}\lambda^P)^{-1}\lambda^{P'}\left\{Y_{1}^{P}-\sum_{j=n+1}^{J}w_jY_{j}^P- \eta^P\left(X_1-\sum_{j=n+1}^{J}w_jX_j\right)\right.\\
&-& \varphi^P(0)\left(M_{1t}^P(0)-\sum_{j=n+1}^{J}w_jM_{jt}^P(0)\right)-\left.\left(\epsilon_{1}^P -\sum_{j=n+1}^{J}w_j\epsilon_{j}^P\right)\right\}.
\end{eqnarray*}
 
Substituting in \ref{diff1} and considering  a generic post-intervention period $t'\geq T$, we have
\begin{eqnarray*}
Y_{1t'}^{0,M_{1}}-\sum_{j=n+1}^{J}w_jY_{jt'}&=&\lambda_{t'}(\lambda^{P'}\lambda^P)^{-1}\lambda^{P'}\left(Y_{1}^{P}-\sum_{j=n+1}^{J}w_jY_{j}^P\right)\\&+&\left(\eta_{t'}-\lambda_{t'}(\lambda^{P'}\lambda^P)^{-1}\lambda^{P'}\eta^P\right)\left(X_1-\sum_{j=n+1}^{J}w_jX_j\right)\\
&-&\lambda_{t'}(\lambda^{P'}\lambda^P)^{-1}\lambda^{P'}\left[\varphi^P(0)(M_1^P(0)-\sum_{j=n+1}^{J}w_jM_{j}^P(0))\right]\\
&+&\varphi_{t'}(0)\left(M_{1t'}(1)-\sum_{j=n+1}^{J}w_jM_{jt'}(0)\right)\\
&-&\lambda_{t'}(\lambda^{P'}\lambda^P)^{-1}\lambda^{P'}\left(\epsilon_1^P-\sum_{j=n+1}^{J}w_j\epsilon_{j}^P\right)+\sum_{j=n+1}^{J}w_j(\epsilon_{1t'}-\epsilon_{jt'}).
\end{eqnarray*}

If we now assume, as we did in the main text, that there exists a set of positive and summing up to 1 weights $W^*$ that satisfies,  $\forall \ t=1,...,T-1$
\begin{eqnarray*}
\sum_{j=n+1}^{J}w_{j}^*Y_{jt}&=& Y_{1t},\\
\sum_{j=n+1}^{J}w_{j}^*X_{j}&=&X_{1},
\end{eqnarray*} 
and $\forall \ t=1,...,T-1, t'$, also satisfies 
$$
\sum_{j=n+1}^{J}w_{j}^*M_{jt}=M_{1t},  
$$
replacing in the post-intervention period, the generic weights  with $W^*$,  we get
\begin{eqnarray*}
Y_{1t'}^{0,M_{1}}-\sum_{j=n+1}^{J}w^*_jY_{jt'}&=& -\lambda_{t'}(\lambda^{P'}\lambda^P)^{-1}\lambda^{P'}\left(\epsilon_1^P-\sum_{j=n+1}^{J}w^*_j\epsilon_{j}^P\right)+\sum_{j=n+1}^{J}w^*_j(\epsilon_{1t'}-\epsilon_{jt'}).
\end{eqnarray*} 

From here, the proof is identical to the one in \cite{Aba2010}. We can write
\begin{eqnarray}
Y_{1t'}^{0,M_{1}}-\sum_{j=n+1}^{J}w_j^*Y_{jt'}=R_{1t'} + R_{2t'}+R_{3t'}\notag 
\end{eqnarray}

where

\begin{eqnarray}
R_{1t'} &=& \lambda_{t'}(\lambda^{P'}\lambda^P)^{-1}\lambda^{P'}\sum_{j=n+1}^{J}w_j^*\epsilon_{j}^P \label{Rs1}\\
R_{2t'} &=& -\lambda_{t'}(\lambda^{P'}\lambda^P)^{-1}\lambda^{P'}\epsilon_1^P \label{Rs2}\\
R_{3t'} &=&  \sum_{j=n+1}^{J}w_j^*(\epsilon_{jt'}-\epsilon_{1t'})\label{Rs3}
\end{eqnarray}
 
Following \cite{Aba2010}, we impose the following assumptions
\begin{assumption}\label{inunit}
$\epsilon_{it}$  $\bot$  $\epsilon_{jt}$ $\forall i \neq j$ with $i, j = 1,...,J.$ 
\end{assumption}
\begin{assumption}\label{intime}
$\epsilon_{it}$  $\bot$  $\epsilon_{it''}$ $\forall t \neq t''$ with $t, t'' = 1,...,t'.$ 
\end{assumption}
\begin{assumption}\label{assmed0}
$E(\epsilon_{it} | X_i, \mu_i, M_{it}(I\{t\geq T\}))=E(\epsilon_{it})=0$ for $i \in \{1, n+1,..., J\}$ and for $t=1,...,t'$
\end{assumption}
Taking the expected value on both sides of \ref{Rs2}, we get 

\begin{eqnarray*}
E(R_{2t'}) &=& E(-\lambda_{t'}(\lambda^{P'}\lambda^P)^{-1}\lambda^{P'}\epsilon_1^P)\\
 &=&-\lambda_{t'}(\lambda^{P'}\lambda^P)^{-1}\lambda^{P'}E(\epsilon_1^P)\\&=&0  
\end{eqnarray*}
 
where the second equality follows from the fact that $-\lambda_{t'}(\lambda^{P'}\lambda^P)^{-1}\lambda^{P'}$ is non-stochastic and the third equality follows from assumption \ref{assmed0}. Taking the expectation on both sides of \ref{Rs3}
\begin{eqnarray*}
E(R_{3t'}) &=&E\left(\sum_{j=n+1}^{J}w_j^*(\epsilon_{jt'}-\epsilon_{1t'})\right)=\sum_{j=n+1}^{J}\left[E(w_j^*\epsilon_{jt'})- E(w_j^*\epsilon_{1t'})\right]\\&=&
\sum_{j=n+1}^{J}\left[E(w_j^*)E(\epsilon_{jt'})- E(w_j^*)E(\epsilon_{1t'})\right]=0
\end{eqnarray*}
where the third equality follows from the fact that weights $W^*=w_{n+1}^*,...,w_J^*$ are determined using constraints on covariates, pre-treatment period outcomes and the mediator, which under assumptions \ref{inunit}, \ref{intime} and \ref{assmed0} are independent from the error terms at time $t' \geq T$. The fourth equality follows from assumption \ref{assmed0}. The remaining \ref{Rs1} can be rewritten as:
\begin{eqnarray} \label{Rs4}
R_{1t'} =\sum_{j=n+1}^{J}w_j^*\sum_{s=1}^{T-1}\lambda_{t'}(\sum_{h=1}^{T-1}\lambda_h'\lambda_h)^{-1}\lambda_s'\epsilon_{js}
\end{eqnarray}
As in \cite{Aba2010}, we further assume that
\begin{assumption} \label{symm}
Let $\varsigma(M)$ be the smallest eigenvalue of 
\begin{eqnarray*}
\frac{1}{M}\sum_{t=T-M+1}^{T-1}\lambda_t'\lambda_t,
\end{eqnarray*}
$\varsigma(M)\geq\underline{\varsigma}>0$ for each positive integer M.
\end{assumption}
\begin{assumption}\label{symm2}
\begin{eqnarray*}
\exists \tx{ }\underline{\lambda} \tx{s.t.} |\lambda_{tf}|\leq\underline{\lambda}\tx{ } \forall \tx{t=1,...,t' and f=1,...,F}.
\end{eqnarray*}
\end{assumption}
Assumption \ref{symm} guarantees that the matrix $\sum_{t=1}^{T}\lambda_t'\lambda_t$ and, consequently, its inverse, are symmetric and positive definite. Thus, for the Cauchy-Schwarz inequality, we find that 
\begin{eqnarray}
\left(\lambda_t\left(\sum_{h=1}^{T-1}\lambda_h'\lambda_h\right)^{-1}\lambda_s'\right)^2 &=& |\langle\lambda_t, A\lambda_s'\rangle|^2 \leq ||A\lambda_t||^2 || A \lambda_s||^2 \label{parent}\\
&=&\left(\lambda_t\left(\sum_{h=1}^{T-1}\lambda_h'\lambda_h\right)^{-1}\lambda_t'\right)\left(\lambda_s\left(\sum_{h=1}^{T-1}\lambda_h'\lambda_h\right)^{-1}\lambda_s'\right) \notag
\end{eqnarray}

Where $A=\left(\sum_{h=1}^{T-1}\lambda_h'\lambda_h\right)^{-1}$. Since $A$ is a symmetric matrix $B=(T-1)A$ is symmetric as well. Thus, it can be decomposed  as $B=GOG^{-1}$. Where $G$ is orthogonal and $G^{-1}=G'$ and $O$ is a diagonal matrix with the eigenvalues of $B$ as elements. Thus,
\begin{eqnarray*}
\lambda_t\left(\sum_{h=1}^{T-1}\lambda_h'\lambda_h\right)^{-1}\lambda_t'=\frac{1}{T-1}(\lambda_tB\lambda_t')=\frac{1}{T-1}(\lambda_tGOG'\lambda_t')
\end{eqnarray*}
Defining $b_t=\lambda_tG$, we have
\begin{eqnarray*}
\lambda_t\left(\sum_{h=1}^{T-1}\lambda_h'\lambda_h\right)^{-1}\lambda_t'=\frac{1}{T-1}(b_tOb_t')=\frac{1}{T-1}\left(b_{t1}^2\frac{1}{\varsigma_1}+\ldots+b_{tF}^2\frac{1}{\varsigma_F}\right)
\end{eqnarray*} 
where $\varsigma_i$ are the eigenvalues of matrix B. From assumption \ref{symm}, imposing $M=T-1$, we'll have that $\frac{1}{\varsigma_i} \leq \frac{1}{\underline{\varsigma}}$ for $i=1,...,F$. Indeed, the eigenvalues of the inverse of a matrix are given by the inverse of the matrix eigenvalues, and B is the inverse of the matrix in assumption \ref{symm}. Consequently:
\begin{eqnarray*}
\lambda_t\left(\sum_{h=1}^{T-1}\lambda_h'\lambda_h\right)^{-1}\lambda_t'&=&\frac{1}{T-1}\sum_{f=1}^{F}\frac{b_{tf}^2}{\varsigma_f} \leq \frac{1}{(T-1)\underline{\varsigma}}\sum_{f=1}^{F}b_{tf}^2\\
&=&\frac{1}{(T-1)\underline{\varsigma}}||b_t||^2=\frac{1}{(T-1)\underline{\varsigma}}||\lambda_tG||^2
\end{eqnarray*}  
As we noticed before, G is an orthogonal and thus isometric matrix, hence $||\lambda_tG||=||\lambda_t||$. Consequently,
\begin{eqnarray*}
\lambda_t\left(\sum_{h=1}^{T-1}\lambda_h'\lambda_h\right)^{-1}\lambda_t' \leq \frac{1}{(T-1)\underline{\varsigma}}||\lambda_t||^2 = \frac{\sum_{f=1}^F\lambda_{tf}^2}{(T-1)\underline{\varsigma}} \leq \frac{\sum_{f=1}^F\underline{\lambda}^2}{(T-1)\underline{\varsigma}}=\frac{F\underline{\lambda}^2}{(T-1)\underline{\varsigma}}
\end{eqnarray*}  
where the last inequality follows from assumption \ref{symm2}. Applying the same idea to the second part of \ref{parent}, we get
\begin{eqnarray}
\left(\lambda_t\left(\sum_{h=1}^{T-1}\lambda_h'\lambda_h\right)^{-1}\lambda_s'\right)^2  &\leq&  \left(\lambda_t\left(\sum_{h=1}^{T-1}\lambda_h'\lambda_h\right)^{-1}\lambda_t'\right)\left(\lambda_s\left(\sum_{h=1}^{T-1}\lambda_h'\lambda_h\right)^{-1}\lambda_s'\right)\notag\\ &\leq& \left(\frac{F\underline{\lambda}^2}{(T-1)\underline{\varsigma}}\label{diseq}\right)^2
\end{eqnarray} 
Following \cite{Aba2010}, we define
\begin{eqnarray}
\overline{\epsilon_j^L}=\sum_{s=1}^{T-1}\lambda_T(\sum_{h=1}^{T-1}\lambda_h'\lambda_h)^{-1}\lambda_s'\epsilon_{js} \label{epsL}
\end{eqnarray} 
for $j=n+1,...,J$. Assume that
\begin{assumption} 
The $p^{th}$ moment of $|\epsilon_{jt}|$ for some even $p$ exists for $j=2,...,J$ and $t=1,...,T-1$
\end{assumption}
Using H\"{o}lder's inequality and taking into account that $0 \leq w_j^* \leq 1$ for $j=n+1,...,J$ we find that:
\begin{eqnarray*}
\sum_{j=n+1}^{J}w_j^*|\overline{\epsilon_j^L}| &=&\sum_{j=n+1}^{J}w_j^*|\overline{\epsilon_j^L}*1| \leq \left(\sum_{j=n+1}^{J}w_j^*|\overline{\epsilon_j^L}|^p\right)^{1/p}\left(\sum_{j=n+1}^{J}w_j^*|1|^q\right)^{1/q}\\
 &=&\left(\sum_{j=n+1}^{J}w_j^*|\overline{\epsilon_j^L}|^p\right)^{1/p}\left(\sum_{j=n+1}^{J}w_j^*\right)^{1/q}=\left(\sum_{j=n+1}^{J}w_j^*|\overline{\epsilon_j^L}|^p\right)^{1/p} \leq \left(\sum_{j=n+1}^{J}|\epsilon_j^L|^p\right)^{(1/p)}
\end{eqnarray*} 
where the last equality follows from $w_{n+1}^*+...+w_J^*=1$ and the last inequality follows from the condition that $w_{n+1}^* \leq 1,..., w_J^* \leq 1$. Applying H\"{o}lder's inequality again, we get
\begin{eqnarray}
E\left[\sum_{j=n+1}^{J}w_j^*|\overline{\epsilon_j^L}|\right] \leq \left(E\left[\sum_{j=n+1}^{J}|\overline{\epsilon_j^L}|^p\right]\right)^{1/p}\label{Hold}
\end{eqnarray} 
Applying Rosenthal's inequality, we have 
\begin{eqnarray*}
E\left[|\overline{\epsilon_j^L}|^p\right] &=& E\left[\left|\sum_{s=1}^{T-1}\lambda_t\left(\sum_{h=1}^{T-1}\lambda_h'\lambda_h\right)^{-1}\lambda_s'\epsilon_{js}\right|\right] \\
&\leq& C\left(p\right) \max\left(\sum_{s=1}^{T-1}E\left[\left|\lambda_t\left(\sum_{h=1}^{T-1}\lambda_h'\lambda_h\right)^{-1}\lambda_s'\epsilon_{js}\right|^p\right]\right.\\&,&\left. \left(\sum_{s=1}^{T-1}E\left[\left|\lambda_t\left(\sum_{h=1}^{T-1}\lambda_h'\lambda_h\right)^{-1}\lambda_s'\epsilon_{js}\right|^2\right]\right)^{p/2}\right)
\end{eqnarray*} 
where $C(p)$ is the $pth$ moment of $-1$ plus a Poisson random variable with mean $1$ (see \cite{Aba2010}). Consider the two elements of $max(.)$. For the first element, we have
\begin{eqnarray*}
\sum_{s=1}^{T-1}E\left[\left|\lambda_t\left(\sum_{h=1}^{T-1}\lambda_h'\lambda_h\right)^{-1}\lambda_s'\epsilon_{js}\right|^p\right]&=&
 \sum_{s=1}^{T-1}E\left[\left(\lambda_t\left(\sum_{h=1}^{T-1}\lambda_h'\lambda_h\right)^{-1}\lambda_s'\right)^{2*\left(p/2\right)}|\epsilon_{js}|^p\right]\\
 &\leq& \sum_{s=1}^{T-1}E\left[\left(\frac{F\underline{\lambda}^2}{\left(T-1\right)\underline{\varsigma}}\right)^{2*\left(p/2\right)}|\epsilon_{js}|^p\right]\\
&=& \left(\frac{F\underline{\lambda}^2}{\underline{\varsigma}}\right)^p\frac{1}{\left(T-1\right)^p}\sum_{s=1}^{T-1}E\left(|\epsilon_{js}|^p\right)
\end{eqnarray*}

where the first equality follows from the distributivity of the power and the inequality follows from \ref{diseq}. For the second element in $max\left(.\right)$, we have
\begin{eqnarray*}
 \left(\sum_{s=1}^{T-1}E\left[\left|\lambda_t\left(\sum_{h=1}^{T-1}\lambda_h'\lambda_h\right)^{-1}\lambda_s'\epsilon_{js}\right|^2\right]\right)^{p/2} 
&\leq& \left[\sum_{s=1}^{T-1}E\left(\left(\frac{F\overline{\lambda}^2}{\left(T-1\right)\underline{\varsigma}}\right)^2\epsilon_{js}^2\right)\right]^{p/2}\\
&=& \left(\frac{F\overline{\lambda}^2}{\underline{\varsigma}}\right)^p\left[\sum_{s=1}^{T-1}\frac{1}{\left(T-1\right)^2}E\left(\epsilon_{js}^2\right)\right]^{p/2}
\end{eqnarray*}
where the first inequality follows from \ref{diseq}. Putting all these results together, we have
\begin{eqnarray*}
E\left[|\overline{\epsilon_j^L}|^p\right]  \leq C\left(p\right)\left(\frac{F\overline{\lambda}^2}{\underline{\varsigma}}\right)^p \max\left(\frac{1}{\left(T-1\right)^p}\sum_{s=1}^{T-1}E\left(|\epsilon_{js}|^p\right), \left[\sum_{s=1}^{T-1}\frac{1}{\left(T-1\right)^2}E\left(\epsilon_{js}^2\right)\right]^{p/2}\right)
\end{eqnarray*} 
As \cite{Aba2010}, we define $\sigma_{js}^2=E|\epsilon_{js}|^2$, $\sigma_j^2 = (1/(T-1)\sum_{s=1}^{T-1}\sigma_{js}^2)$, $\overline{\sigma^2}=max_{j=n+1,...,J}\sigma_j^2$ and $\overline{\sigma}=\sqrt{\overline{\sigma^2}}$. Similarly, we define $\tau_{p,jt}=E|\epsilon_{jt}|^p$, $\tau_{p,j}=\frac{1}{(T-1)}\sum_{t=1}^{T-1}\tau_{p, jt}$, and $\overline{\tau_p}=max_{j=n+1,...,J}\tau_{p,j}$. We can write the first element of  $max(.)$ as
\begin{eqnarray*}
\frac{1}{(T-1)^p}\sum_{s=1}^{T-1}E(|\epsilon_{js}|^p) = \frac{1}{(T-1)^{p-1}}\frac{1}{(T-1)}\sum_{t=1}^{T-1}\tau_{pjt}=\frac{1}{(T-1)^{p-1}}\tau_{pj}
\end{eqnarray*} 
Similarly, the second element can be written as
\begin{eqnarray*}
\left[\sum_{s=1}^{T-1}\frac{1}{(T-1)^2}E(\epsilon_{js}^2)\right]^{p/2}=\left(\frac{1}{T-1}\frac{1}{T-1}\sum_{s=1}^{T-1}\sigma_{js}^2\right)^{p/2}=\left(\frac{1}{T-1}\sigma_j^2\right)^{p/2}
\end{eqnarray*} 

Thus, defining $\varpi=C(p)(\frac{F\overline{\lambda}^2}{\underline{\varsigma}})^p$, we have
 \begin{eqnarray*}
E\left[|\overline{\epsilon_j^L}|^p\right]  &\leq& \varpi \max\left(\frac{1}{\left(T-1\right)^{p-1}}\tau_{pj}, \left(\frac{1}{T-1}\sigma_j^2\right)^{p/2}\right)\\
\sum_{j=n+1}^{J}E\left[|\overline{\epsilon_j^L}|^p\right] &=& E\left[\sum_{j=n+1}^{J}|\overline{\epsilon_j^L}|^p\right]\\
 &\leq& \varpi \max\left(\frac{1}{\left(T-1\right)^{p-1}}\sum_{j=n+1}^{J}\tau_{pj}, \sum_{j=n+1}^{J}\left(\frac{1}{T-1}\sigma_j^2\right)^{p/2}\right) \\&=& \varpi \max\left(\frac{J-n-1}{\left(T-1\right)^{p-1}}\frac{1}{J-n-1}\sum_{j=n+1}^{J}\tau_{pj}, \frac{1}{\left(T-1\right)^{p/2}}\sum_{j=n+1}^{J}\sigma_j^{2*p/2}\right) \\
\left(E\left[\sum_{j=n+1}^{J}|\overline{\epsilon_j^L}|^p\right]\right)^{1/p} &\leq& \varpi^{1/p} \max\left(\frac{\left(\frac{J-n-1}{\left(T-1\right)^{p-1}}\right)^{1/p}}{\left(J-n-1\right)^{1/p}}\left(\sum_{j=n+1}^{J}\tau_{pj}\right)^{1/p}, \frac{\left(\sum_{j=n+1}^{J}\sigma_j^{2*p/2}\right)^{1/p}}{\left(T-1\right)^{\left(p/2\right)*\left(1/p\right)}}\right) 
\\&=& \varpi^{1/p} \max\left(\left(\frac{J-n-1}{\left(T-1\right)^{p-1}}\right)^{1/p}\overline{\tau}_{p}^{1/p}, \frac{1}{\left(T-1\right)^{1/2}}\left(\sum_{j=n+1}^{J}\overline{\sigma}^{2*\left(p/2\right)}\right)^{1/p}\right)
\end{eqnarray*}
where the last equality follows from $\frac{1}{J-n-1}\sum_{j=n+1}^{J}\tau_{pj}=E(\tau_{pj}) \leq max_{j}(\tau_{pj})=\overline{\tau_p}$. Thus,
\begin{eqnarray}
\left(E\left[\sum_{j=n+1}^{J}|\overline{\epsilon_j^L}|^p\right]\right)^{1/p} &\leq& \varpi^{1/p} max\left(\frac{\left(J-n-1\right)^{1/p}\overline{\tau}_{p}^{1/p}}{\left(T-1\right)^{1-1/p}}, \frac{\left(J-n-1\right)\overline{\sigma}^{2*\left(p/2\right)}}{\left(T-1\right)^{1/2}}\right)^{1/p}\notag\\
&=&\varpi^{1/p}\left(J-n-1\right)^{1/p}max\left(\frac{\overline{\tau}_{1/p}}{\left(T-1\right)^{1-\frac{1}{p}}}, \frac{\sqrt{\overline{\sigma}^2}}{\left(T-1\right)^{1/2}}\right)\label{esumel}
\end{eqnarray}
this implies
\begin{eqnarray*}
E\left[|R_{1t'}|\right]&=&E\left[\left|\sum_{j=n+1}^{J}w_j^*\epsilon_j^L\right|\right]\\
&\leq& E\left[\sum_{j=n+1}^{J}w_j^*|\epsilon_j^L|\right] \\&\leq& \left(E\left[\sum_{j=n+1}^{J}|\epsilon_j^L|^p\right]\right)^{1/p} \\&\leq& \varpi^{1/p} (J-n-1)^{1/p}max\left(\frac{\overline{\tau}_p^{1/p}}{\left(T-1\right)^{1-\frac{1}{p}}}, \frac{\overline{\sigma}}{\left(T-1\right)^{1/2}}\right)
\end{eqnarray*} 
where, in the second equation, the first equality follows from \ref{Rs2} and \ref{epsL}, the first inequality follows from the triangular inequality, the second follows from \ref{Hold} and the third from \ref{esumel}. It follows that
\begin{eqnarray*}
E|R_{1t'}| \leq C(p)^{1/p}\frac{\overline{\lambda^2}F}{\underline{\varsigma}}(J-n-1)^{1/p}\max\left\{\frac{\overline{\tau_p^{1/p}}}{(T-1)^{1-1/p}}, \frac{\overline{\sigma}}{(T-1)^{1/2}}\right\}.
\end{eqnarray*}
Thus, the difference between the expected value of $Y_{1t}^{0,M_{1}}$ and its synthetic counterpart can be bound by something that goes to zero when the number of pre-intervention periods goes to infinity, namely 
\begin{eqnarray*}
E\left(Y_{1t'}^{0,M_{1}}-\sum_{j=n+1}^{J}w_j^*Y_{jt'}\right)=E(R_{1t'})=o(T).
\end{eqnarray*}
\section{Identification of $\delta_{i{t'}}(1)$}\label{indirect}
Finding a ``synthetic'' value of $Y_{1t}^{1, M_{0}}$ is more challenging and requires more than 1 treated unit. First, we need to estimate what  value  the mediator of unit $1$ would have taken in the absence of the intervention ($M_{1t}(0)$). This could be done with a standard SCM, using the mediator as an outcome. Second, we propose to treat the remaining treated units as control units in a SCM where we also use the distance between the first step estimate of $M_{1t}(0)$ and the other treated units mediators, in computing the weights.   
If the number of treated units is big enough, we can also create a ``synthetic'' $Y_{i{t'}}^{1,M_{0}}$. This is done in two steps.
In a first step, we estimate $M_{1{t'}}(0)$ by $\hat{M}_{1{t'}}(0)=\sum_{i=n+1}^{J}k_{i{t'}}^*M_{i{t'}}$ with $K_{{t'}}^*=(k_{n+1,{t'}}^*,\ldots,k_{J{t'}}^*)$ chosen with a standard SCM. Note that also those weights need to be calculated for each ${t'}$. In a second step, we need to find a vector of positive and adding up to 1 weights $Q_{{t'}}^*=(q_{2{t'}}^*,...,q_{n{t'}}^*)$, such that $Y_{i{t'}}^{1,M_{0}}=\sum_{i=2}^n q^*_{i{t'}}Y_{i{t'}}$. $Q^*_{{t'}}$ is estimated with a SCM but using only the other treated units. More specifically, let $\Omega^{\delta_{t'}(1)}_1=(X_1, Y_{11},\ldots,Y_{1,T-1},M_{11},\ldots,M_{1,T-1}, \hat{M}_{1{t'}}(0))$, $\omega^{\theta_{t'}(1)}_{0i}=(X_i, Y_{11},\ldots,Y_{i,T-1},M_{11}$, $\ldots,M_{i,T-1},M_{i,{t'}})$, and $\Omega^{\theta_{t'}(1)}_{0}=(\omega^{\theta_{t'}(1)}_{2},\ldots,\omega^{\theta_{t'}(1)}_{n})'$, then
\begin{eqnarray*}
Q_{{t'}}^*&=&\min_{q_{n+1,{t'}},...,q_{J{t'}}} ||\Omega^{\theta_{t'}(1)}_{1}-Q_{{t'}} \Omega^{\theta_{t'}(1)}_{0}||_V\\
&s.t.& q_{n+1,{t'}}\leq 0,...,q_{J{t'}}\leq 0, \sum_{i=n+1}^J q_{i{t'}}=1,
\end{eqnarray*}
where the distance and $V$ are defined as above for $Y_{i{t'}}^{0,M_{1}}$.

Let $\hat{Y}_{1{t'}}^{1,M_{0}}=\sum_{i=2}^n q^*_{i{t'}}Y_{i{t'}}$, similar as before, we assume that $Q^*_{{t'}}$ exists and satisfies $\forall \ t=1,...,T-1$
\begin{eqnarray*}
\sum_{j=2}^{n}q_{j{t'}}^*Y_{jt}&=& Y_{1t},\\
\sum_{j=2}^{n}q_{j{t'}}^*X_{j}&=&X_{1},\\
\sum_{j=2}^{n}q_{j{t'}}^*M_{jt}=M_{1t},
\end{eqnarray*} 

$\forall \ t=1,...,T-1$
and  
$$
\sum_{j=2}^{n}q_{j{t'}}^*M_{j{t'}}=\hat{M}_{1{t'}}(0).
$$
Under extra standard conditions and assuming that $\rho_{t'}(\cdot)$ is a linear function
$$ 
E(\hat{Y}_{1{t'}}^{1,0})=Y_{1{t'}}^{1,M_{0}}+o(T).
$$

The latter assumption can admittedly be restrictive in many applications. However, it is substantially weaker than assuming a constant $\rho_{t'}$.
Then, we can estimate  the indirect effect $\delta_{i{t'}}(1)$ and  the direct effect as ${\theta}_{1{t'}}(M_{1t}(0))$ as $$\hat{\delta}_{1{t'}}(1)=Y_{1{t'}}-\hat{Y}_{1{t'}}^{1,M_{0}}, \qquad \hat{\theta}_{1t'}(M_{1t'}(0))=\hat{\alpha}_{i{t'}}-\hat{\delta}_{1{t'}}(1),$$ 
respectively. Intuitively, $Q^*_{{t'}}$ exists under the similar assumptions as the one discussed in the main text. However, if the number of treated units is too small $\hat{Y}_{1{t'}}^{1,M_{0}}$ will be a very poor approximation of $Y_{1{t'}}^{1,M_{0}}$. In this setting, it is only possible to estimate $\delta_{i{t'}}(0)$ and $\theta_{i{t'}}(1)$. 

\section{Extra assumptions on the mediator needed for $Y_{1t}^{1M_{0}}$}\label{appAss}
To create a synthetic $Y_{1t}^{1M_{0}}$, we need to impose the standard SCM assumptions on the mediator, which are: 
\begin{assumption}
$\sum_{t=1}^{T-1}\vartheta_t'\vartheta_t$ is non-singular. \par
\end{assumption}
\begin{assumption}
$\nu_{it}$  $\bot$  $\nu_{jt}$ $\forall i \neq j$ with $i, j \in \{1,n+1,...,J\}.$ 
\end{assumption}
\begin{assumption}
$\nu_{it}$  $\bot$  $\nu_{it''}$ $\forall t \neq t''$ with $t, t'' = 1,...,t'.$ 
\end{assumption}
\begin{assumption}
$E(\nu_{it} | \{Z_i, \varrho_i\}_{i\in\{1, n+1,...,J\}})=E(\nu_{it})=0$ for $i \in \{1, n+1,..., J\}$ and for $t=1,...,t'$
\end{assumption}
\begin{assumption}
$\kappa(M)\geq\underline{\kappa}>0$ for each positive integer M, where $\kappa(M)$ is the smallest eigenvalue of 
\begin{equation}
\frac{1}{M}\sum_{t=T-M+1}^{T-1}\vartheta_t'\vartheta_t.
\end{equation}
\end{assumption}
\begin{assumption}
\begin{equation}
\exists \tx{ }\underline{\vartheta} \tx{s.t.} |\vartheta_{tv}|\leq\underline{\vartheta} \tx{ }\forall \tx{t=1,...,t' and v=1,...,V}.
\end{equation}
\end{assumption}
\begin{assumption}
$\exists$ a $p^{th}$ moment of $|\nu_{jt}|$ for some even p and for $j=n+1,...,J$ and $t=1,...,t'$ \par
\end{assumption}

\section{Derivation of ``Synthetic'' $Y_{1t}^{1M_{0}}$}\label{appY10}
As for $Y_{1t}^{0M_{1}}$, we drop the subscript $t$ from the weight and write
\begin{eqnarray*}
\sum_{j=2}^{n}q_jY_{jt}=\zeta_t &+& \eta_t\sum_{j=2}^{n}q_jX_j+\lambda_t\sum_{j=2}^{n}q_j\mu_j+\varphi_t\left(I\{t\geq T\}\right)\sum_{j=2}^{n}q_jM_{jt}\left(I\{t\geq T\}\right)\\&+&\sum_{j=2}^{n}q_j\rho_t\left(M_{jt}\left(I\{t\geq T\}\right)\right)I\{t\geq T\}+\sum_{j=2}^{n}q_j\epsilon_{jt}.
\end{eqnarray*}
Thus,
\begin{eqnarray*}
Y_{1t}^{1,M_{0}}-\sum_{j=2}^{n}q_jY_{jt}&=&\eta_t\left(X_1-\sum_{j=2}^{n}q_jX_j\right)+\lambda_t\left(\mu_1-\sum_{j=2}^{n}q_j\mu_j\right)\\
&+&\varphi_t\left(I\{t\geq T\}\right)\left(M_{1t}(0)-\sum_{j=2}^{n}q_j M_{jt}(I\{t\geq T\})\right)\\
&+&\left(\rho_t\left(M_{1t}\left(0\right)\right)-\sum_{j=2}^{n}q_j\rho_t\left(M_{jt}\left(I\{t\geq T\}\right)\right)\right)I\{t\geq T\}\\
&+&\sum_{j=2}^{n}q_j\left(\epsilon_{1t}-\epsilon_{jt}\right)
\end{eqnarray*}
Using the same notation as before in the pre-intervention period, we have
\begin{eqnarray*}
Y_{1}^{P}-\sum_{j=2}^{n}q_jY_{j}^P &=&\eta^P\left(X_1-\sum_{j=2}^{n}q_jX_j\right)+ \lambda^P\left(\mu_1-\sum_{j=2}^{n}q_j\mu_j\right)\\&+& \varphi^P\left(0\right)\left(M_{1}^P\left(0\right)-\sum_{j=2}^{n}q_jM_{j}^P\left(0\right)\right)+\left(\epsilon_{1}^P-\sum_{j=2}^{n}q_j\epsilon_{j}^P\right)\label{diff2}
\end{eqnarray*}
Thus,
\begin{eqnarray*}
\lambda^P\left(\mu_1-\sum_{j=2}^{n}q_j\mu_j\right) &=&Y_{1}^{P}-\sum_{j=2}^{n}q_jY_{j}^P - \eta^P\left(X_1-\sum_{j=2}^{n}q_jX_j\right)\\&-&
\varphi^P\left(0\right)\left(M_{1}^P\left(0\right)-\sum_{j=2}^{n}q_jM_{j}^P\left(0\right)\right)-\left(\epsilon_{1}^P-\sum_{j=2}^{n}q_j\epsilon_{j}^P\right)
\end{eqnarray*}
Multiplying both sides by $(\lambda^{P'}\lambda^P)^{-1}\lambda^{P'}$, we get
\begin{eqnarray*}
\mu_1-\sum_{j=2}^{n}q_j\mu_j &=&\left(\lambda^{P'}\lambda^P\right)^{-1}\lambda^{P'}\left\{\left(Y_{1}^{P}-\sum_{j=2}^{n}q_jY_{j}^P\right)-\eta^P\left(X_1-\sum_{j=2}^{n}q_jX_j\right)\right.\\&-&\left.\varphi^P\left(0\right)\left(M_{1}^P\left(0\right)-\sum_{j=2}^{n}q_jM_{j}^P\left(0\right)\right)-\left(\epsilon_{1}^P-\sum_{j=2}^{n}q_j\epsilon_{j}^P\right)\right\}.
\end{eqnarray*}

Substituting in \ref{diff2} and considering a generic post-intervention period t', we have
\begin{eqnarray*}
Y_{1t'}^{1,M_{0}}-\sum_{j=2}^{n}q_jY_{jt'}&=&\left(\lambda^{P'}\lambda^P\right)^{-1}\lambda^{P'}\left(Y_{1}^{P}-\sum_{j=2}^{n}q_jY_{j}^P\right)\\&+&
\left(\eta_{t'}-\left(\lambda^{P'}\lambda^P\right)^{-1}\lambda^{P'}\eta^P\right)\left(X_1-\sum_{j=2}^{n}q_jX_j\right)\\&-&\left(\lambda^{P'}\lambda^P\right)^{-1}\lambda^{P'}\varphi^P\left(0\right)\left(M_{1}^P\left(0\right)-\sum_{j=2}^{n}q_jM_{j}^P\left(0\right)\right)\\&+&\varphi_{t'}(1)\left(M_{1t'}(0)-\sum_{j=2}^{n}q_j M_{jt'}(1)\right)\\
&+&\left(\rho_{t'}\left(M_{1t'}\left(0\right)\right)-\sum_{j=2}^{n}q_j\rho_{t'}\left(M_{jt}\left(1\right)\right)\right)\\
&-&
\left(\lambda^{P'}\lambda^P\right)^{-1}\lambda^{P'}\left(\epsilon_{1}^P-\sum_{j=2}^{n}q_j\epsilon_{j}^P\right)+\sum_{j=2}^{n}q_j\left(\epsilon_{1t'}-\epsilon_{jt'}\right)
\end{eqnarray*}
Assume, as we did in the main text, that weights $q_2^*,...,q_n^*$ exists that satisfy $\forall t=1,...,T-1$
\begin{eqnarray*}
\sum_{j=2}^{n}q_{j}^*Y_{jt}&=& Y_{1t},\\
\sum_{j=2}^{n}q_{j}^*X_{j}&=&X_{1}, \\
\sum_{j=2}^{n}q_{j}^*M_{jt}&=&M_{1t},
\end{eqnarray*} 
and it also satisfies 
$$
\sum_{j=2}^{n}q_{j}^*M_{jt'}=\hat{M}_{1t'}(0).  
$$ 
Substituting the generic weights with $q_2^*,...,q_n^*$ in the  post-intervention period $t'$, we get
\begin{eqnarray*}
Y_{1t'}^{1,M_{0}}-\sum_{j=2}^{n}q_j^*Y_{jt'}
&=&\left(\rho_{t'}\left(M_{1t'}\left(0\right)\right)-\sum_{j=2}^{n}q^*_j\rho_{t'}\left(M_{jt'}\left(1\right)\right)\right)\\
&-&
\left(\lambda^{P'}\lambda^P\right)^{-1}\lambda^{P'}\left(\epsilon_{1}^P-\sum_{j=2}^{n}q^*_j\epsilon_{j}^P\right)+\sum_{j=2}^{n}q^*_j\left(\epsilon_{1t'}-\epsilon_{jt'}\right)
\end{eqnarray*}
Note that, as by assumption $\sum_{j=2}^{n}q_{j}^*M_{t'}=\hat{M}_{1t'}(0)$, and $\hat{M}_{1t'}(0)$ is estimated using a standard SCM 
\begin{eqnarray*}
E\left(\varphi_{t'}\left(1\right)\left(M_{1t'}(0)-\sum_{j=2}^{n}q^*_j M_{jt'}(1)\right)\right)=o(T).
\end{eqnarray*}

As mentioned earlier, for identification we have to impose an extra assumption, namely:
\begin{assumption}\label{dirlin}
$\rho_{t'}(.)$ is a linear function
\end{assumption}
Under assumption \ref{dirlin}, we have
\begin{eqnarray*}
E\left[\left(\rho_{t'}\left(M_{1t'}\left(0\right)\right)-\sum_{j=2}^{n}q^*_j\rho_{t'}\left(M_{jt'}\left(1\right)\right)\right)\right]&=& E\left[\left(\rho_{t'}\left(M_{1t'}\left(0\right)\right)-\rho_{t'}\left(\sum_{j=2}^{n}q^*_j M_{jt'}\left(1\right)\right)\right)\right],\\
&=&E\left[\left(\rho_{t'}\left(M_{1t'}\left(0\right)\right)-\rho_{t'}\left(\hat{M}_{1t'}(0)\right)\right)\right],\\
&=&\rho_{t'}\left(M_{1t'}\left(0\right)\right)-\rho_{t'}\left(E(\hat{M}_{1t'}(0))\right)=o(T).
\end{eqnarray*}
Thus, 
\begin{eqnarray*}
Y_{1t'}^{1,M_{0}}-\sum_{j=2}^{n}q_j^*Y_{jt'}&=&-\left(\lambda^{P'}\lambda^P\right)^{-1}\lambda^{P'}\left(\epsilon_{1}^P-\sum_{j=2}^{n}q^*_j\epsilon_{j}^P\right)+\sum_{j=2}^{n}q^*_j\left(\epsilon_{1t'}-\epsilon_{jt'}\right).
\end{eqnarray*} 

This can be shown to imply 
\begin{eqnarray*}
E(Y_{1t'}^{1,M_{0}}-\sum_{j=2}^{n}q_j^*Y_{jt'})&=&o(T).
\end{eqnarray*}

\section{Data Sources and Donor Pool}\label{data}
The variables employed stem from different sources. As mentioned in \ref{set}, data on the outcome and on the mediator stem from the Tax Burden on Tobacco compilation (\citealt{Orz2005}). We obtained the data on population from the US Census Bureau. We obtained the data on the real GDP per capita, chained at 1997 prices, from the Bureau of Economic Analysis of the US Department of Commerce. Finally, we obtained the data on beer consumption per capita from \cite{Kap2018}. \\

In Table \ref{tab1}, we present the donor pool for the total (first two columns) and for the direct (third and fourth columns) effect estimations with the corresponding weights. In the direct effect estimation, different weights are used in each post-treatment period. For simplicity, we report those found in the last post-treatment period.  

\begin{table}[H]
\centering
\caption{Donor Pool}\label{tab1}
\begin{tabular}{lccc} \hline
Donor Pool & Weights Total Effect & Weights Direct Effects \\ \hline
Alaska & - & 0.046 \\
Alabama & <0.001 & <0.001 \\
Arkansas & <0.001 & <0.001 \\
Colorado & <0.001 & <0.001 \\
Connecticut & 0.014 & <0.001 \\
Delaware & <0.001 & <0.001 \\
Georgia & <0.001 & <0.001 \\
Hawaii & - & 0.135 \\
Iowa & <0.001 & <0.001 \\
Idaho & 0.005 & <0.001 \\
Illinois & <0.001 & <0.001 \\
Indiana & <0.001 & <0.001 \\
Kansas & <0.001 & <0.001 \\
Kentucky & <0.001 & <0.001 \\
Lousiana & <0.001 & <0.001 \\
Maryland & - & 0.001 \\
Maine & <0.001 & <0.001 \\
Michigan & - & <0.001 \\
Minnesota & <0.001 & <0.001 \\
Missouri & <0.001 & <0.001 \\
Mississippi & <0.001 & <0.001 \\
Montana & 0.061 & <0.001 \\
North Carolina & 0.005 & <0.001 \\
North Dakota & <0.001 & <0.001 \\
Nebraska & <0.001 & <0.001 \\
New Hampshire & 0.107 & <0.001 \\
New Jersey & - & 0.001 \\
New Mexico & 0.349 & <0.001 \\
Nevada & 0.032 & 0.263 \\
New York & - & 0.249 \\
Ohio & <0.001 & <0.001 \\
Oklahoma & <0.001 & <0.001 \\
Pennsylvania & <0.001 & <0.001 \\
Rhode Island & 0.006 & <0.001 \\
South Carolina & <0.001 & <0.001 \\
South Dakota & <0.001 & <0.001 \\
Tennessee & <0.001 & <0.001 \\
Texas & 0.001 & <0.001 \\
Utah & 0.247 & 0.302 \\
Virginia & <0.001 & <0.001 \\
Vermont & <0.001 & <0.001 \\
Washington & - & <0.001 \\
Wisconsin & 0.001 & <0.001 \\
West Virginia & 0.169 & <0.001 \\
Wyoming & <0.001 & <0.001 \\
\hline
\end{tabular}
\end{table}
\section{Mediator Fits}\label{med_con}
\begin{figure}[H]
\begin{subfigure}{0.5\textwidth}
\centering\captionsetup{width=.8\linewidth}
\includegraphics[width=0.9\linewidth, height=7cm]{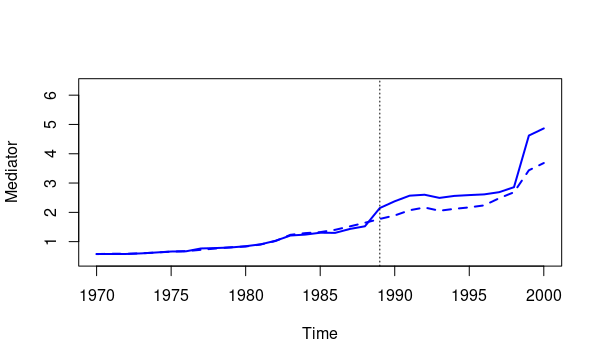}
\caption{Total Effect}
\label{fig2}
\end{subfigure}
\begin{subfigure}{0.5\textwidth}
\centering\captionsetup{width=.8\linewidth}
\includegraphics[width=0.9\linewidth, height=7cm]{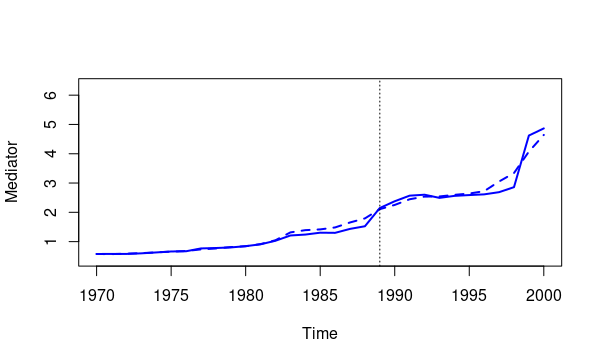}
\caption{Direct Effect}
\label{fig4}
\end{subfigure}
\caption{Constraints on the mediator for the estimation of the total (left) and those of the direct (right) effects. We are mostly interested in post-treatment overlaps in the right panel. The solid line represents the treated unit mediator. The dashed line represents the synthetic unit mediator. The vertical line corresponds to the first treatment period. In the direct effect estimation, different weights are used in each post-treatment period. For simplicity, we use those found in the last post-treatment period.}
\label{med1}
	\end{figure}

\section{Spillover Effects Testing and Leave-one-out Robustness Checks} \label{spill}
\begin{figure}[H]
\begin{subfigure}{0.5\textwidth}
\centering\captionsetup{width=.8\linewidth}
\includegraphics[width=0.9\linewidth, height=7cm]{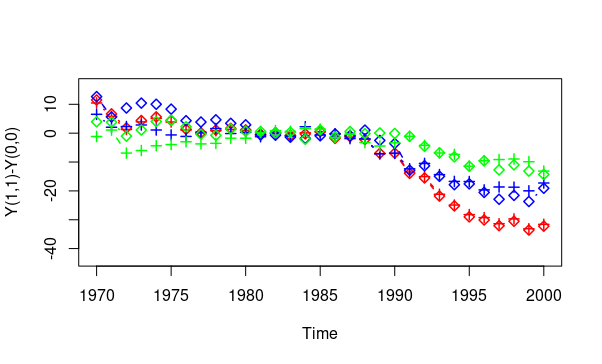} 
\caption{Sensitivity of the results to the exclusion, from the donor pool, of the state of Nevada. Total, direct, and indirect effects are represented, respectively, in red, blue and green. The diamond-line represents baseline estimation. The x-line represents the spillover-free estimation.}
\label{spill1}
\end{subfigure}
\begin{subfigure}{0.5\textwidth}
\centering\captionsetup{width=.8\linewidth}
\includegraphics[width=0.9\linewidth, height=7cm]{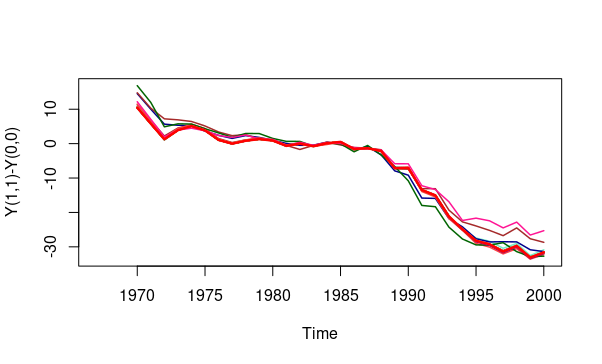} 
\caption{Leave-one-out robustness checks for the total effect. The thick red line represents the baseline estimation.}
\label{loo_tot}
\end{subfigure}
\caption{Robustness checks}
\end{figure}

\begin{figure}[h!]
\begin{subfigure}{0.5\textwidth}
\centering\captionsetup{width=.8\linewidth}
\includegraphics[width=0.9\linewidth, height=7cm]{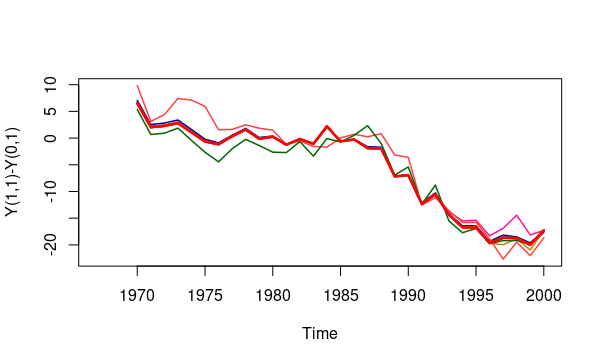} 
\caption{Leave-one-out robustness checks for the direct effect. The thick red line represents the baseline estimation.}
\label{loo_dir}
\end{subfigure}
\begin{subfigure}{0.5\textwidth}
\centering\captionsetup{width=.8\linewidth}
\includegraphics[width=0.9\linewidth, height=7cm]{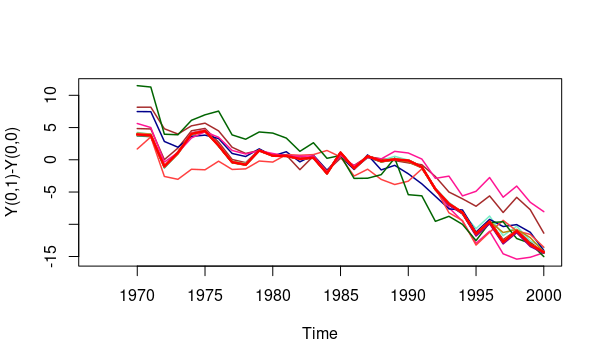} 
\caption{Leave-one-out robustness checks for the indirect effect. The thick red line represents the baseline estimation.}
\label{loo_ind}
\end{subfigure}
\caption{Robustness checks}
\end{figure}

\newpage \setlength\baselineskip{14.0pt}
\bibliographystyle{ecta}
\bibliography{bibliographySCM}

\end{document}